\documentclass[fleqn,usenatbib]{mnras}

\usepackage{newtxtext,newtxmath}

\usepackage[T1]{fontenc}

\DeclareRobustCommand{\VAN}[3]{#2}
\let\VANthebibliography\thebibliography
\def\thebibliography{\DeclareRobustCommand{\VAN}[3]{##3}\VANthebibliography}

\usepackage{graphicx}	
\usepackage{amsmath}	
\usepackage{xcolor}
\usepackage{multirow}
\usepackage{array}
\usepackage{comment}

\usepackage{pgfplots}

\newcommand{\mkpink}{\color{pink}}

    \usetikzlibrary{intersections}
    \pgfplotsset{compat=1.11}
\pgfplotsset{
    every axis plot/.append style = {font = \normalsize}
  }
  
\newcolumntype{x}[1]{wc{#1}}
\newcolumntype{M}[1]{>{\centering\arraybackslash}m{#1}}

\newcommand{\mz}[1]{\textcolor{magenta}{MZ: #1}}

\title[Radiative Cooling Effects on Horizon Polarization]{From Morphology to Variability: Radiative Cooling Effects on Horizon-Scale Polarization in Two-Temperature GRMHD Simulations}

\author[Long et al.]
{Jane SiNan Long$^{1,2}$\thanks{E-mail: sinan.long.23@ucl.ac.uk (JSL), akhil\_uniyal@sjtu.edu.cn (AU), mzhang22@sjtu.edu.cn (MZ), mizuno@sjtu.edu.cn (YM)}, Akhil Uniyal$^{2}$, Mingyuan Zhang$^{2}$,  Yosuke Mizuno$^{2,3,4,5}$, Ziri Younsi$^{1}$, 
\and Christian M. Fromm$^{6,5}$
and Kinwah Wu$^{1,7,8}$   
\\
$^{1}$Mullard Space Science Laboratory, University College London, 
  Holmbury St Mary, Surrey, RH5 6NT, United Kingdom \\
  $^{2}$Tsung-Dao Lee Institute, Shanghai Jiao Tong University, Shanghai 201210, China \\
  $^{3}$School of Physics and Astronomy, Shanghai Jiao Tong University, Shanghai 200240, China \\  
  $^{4}$Key Laboratory for Particle Astrophysics and Cosmology (MOE) and Shanghai Key Laboratory for Particle Physics and Cosmology, \\ 
  \hspace*{0.5em}Shanghai Jiao Tong University, Shanghai 200240, China \\
  $^{5}$Institut f\"ur Theoretische Physik, Goethe-Universit\"at Frankfurt, D-60438 Frankfurt am Main, Germany \\
   $^{6}$Institut f\"ur Theoretische Physik und Astrophysik, Universit\"at W\"urzburg, D-97074 W\"urzburg, Germany \\
   $^{7}$Department of Physics, 
    Chinese University of Hong Kong, Shatin, Hong Kong SAR, China \\ 
  $^{8}$Kavli Institute for the Physics and Mathematics of the Universe (WPI), UTIAS, The University of Tokyo, Kashiwa, Chiba 277-8583, Japan
}

\date{Accepted XXX. Received YYY; in original form ZZZ}

\pubyear{\the\year{}}

\begin{document}
\label{firstpage}
\pagerange{\pageref{firstpage}--\pageref{lastpage}}
\maketitle

\begin{abstract}


Polarization signatures provide a new window to investigate the effects of radiative cooling in the horizon-scale accretion flows. Morphology and variability of polarization offer quantifiable diagnostics of how cooling modifies the polarised emission from two-temperature GRMHD simulations.
We find that cooling enhances the effective Faraday depth, leading to stronger large-scale Faraday scrambling, particularly at higher accretion rates. In contrast, depolarization associated with higher-order photons is comparable between cooling and non-cooling models.
Radiative cooling also increases the intrinsic asymmetry in both the ring structure and the polarization pattern. This effect is quantified by enhanced power in non-axisymmetric azimuthal modes ($\beta_m$, $m \neq 2$) relative to the dominant quadrupolar component $\beta_2$. The increased asymmetry is directly linked to stronger temporal variability of the polarization angle $\angle\beta_2$, including frequent sign reversals that are absent in non-cooling models.
The radial profile of $\angle \beta_2$ further localizes the physical origin of these effects, distinguishing regions dominated by Faraday rotation from those influenced by photon ring contributions, and providing a clear separation between cooling and non-cooling cases. Additional tests including a non-thermal electron population indicate that the polarization structure at 230 GHz is largely insensitive to the detailed form of the electron distribution functions.
Our results demonstrate that horizon-scale polarization asymmetry, variability, and radial structure encode robust signatures of radiative cooling. These findings highlight the diagnostic power of time-resolved polarimetry and high-resolution imaging for constraining radiative processes in black hole accretion flows with EHT-like observations.


\end{abstract}




\section{Introduction}

Accretion flows around low-luminosity supermassive black holes are hot, low-density, and weakly collisional plasmas, indicating that electrons and ions do not efficiently equilibrate via Coulomb collisions \citep[e.g.,][]{Yuan:2014gma}. Consequently, the electron-to-ion temperature ratio is regulated by collisionless kinetic processes such as turbulent dissipation, magnetic reconnection, and wave–particle interactions. 
Radiative cooling processes, such as bremsstrahlung, synchrotron radiation of the thermal electrons \citep{Esin1996ApJ}, and multiple inverse Compton scattering of the synchrotron photons by the thermal electrons \citep{Narayan1995ApJ}, also play an important role in the evolution of the accretion flows. Determining how electrons heat up from the ions and their energy dissipation is therefore a central problem in plasma physics, with relevance ranging from the solar wind to laboratory experiments. In the context of horizon-scale black hole accretion flows, empowered by the unprecedented constraints provided by recent Event Horizon Telescope (EHT) observations of M87* \citep{2019ApJ...875L...1E} and Sgr~A* \citep{2022ApJ...930L..12E}, General-relativistic magnetohydrodynamic (GRMHD) simulations provide a realistic time-dependent, turbulent nature of the accretion flow.

Standard single-fluid GRMHD simulations implicitly assume that electrons and ions are perfectly co-moving and thermally coupled. In contrast, two-temperature prescriptions allow the electron temperature to be evolved alongside the ion temperature \citep{Ressler2015MNRAS,Mizuno2021MNRAS}, producing sufficiently hot electrons to explain hard X-ray emission in sources such as Cyg X-1 \citep{Shapiro1976ApJ}, as well as the spectral properties of X-ray binaries and active galactic nuclei \citep{Ichimaru1977ApJ, Rees1982Natur, Wandel1991ApJ, Luo1994MNRAS}. A widely used empirical approach is the $R$–$\beta$ prescription \citep{Mocibrodzka2016A&A}, which parameterizes the electron temperature as a function of the plasma beta. This model has been extensively employed in GRMHD-based interpretations of EHT observations of M87* and Sgr~A* \citep{EHT2019ApJ...875L...5E, EHT2021ApJ...910L..13E, 2022ApJ...930L..16E}. However, such prescriptions lack direct physical grounding, making it difficult to extract robust conclusions about the underlying electron heating mechanisms.

An alternative, physics-motivated approach evolves the electron temperature self-consistently by solving an additional electron entropy equation. In this framework, physically motivated heating channels such as turbulent dissipation and magnetic reconnection are explicitly modeled \citep{Ressler2015MNRAS, Ressler2017MNRAS}. 
 To obtain complete information on electron thermodynamics, electron feedback on the fluid through Coulomb coupling and radiative processes \citep{Sadowski2017MNRAS} are incorporated inside the GRMHD simulations. More recently, \cite{Dihingia2023MNRAS} combined these elements into a unified electron-entropy evolution scheme that accounts for both physical heating mechanisms and electron thermodynamics using a self-consistent two-temperature paradigm.


In this paper, we investigate the polarization properties of near-horizon accretion flows. To do it, we use the 3D two-temperature GRMHD simulations of magnetized accretion flows onto a highly rotating black hole with radiative cooling \citep{2026arXiv260515502Z}. 
polarised emission provides a powerful diagnostic of the physical conditions in the innermost regions of accretion flows, as it is directly sensitive to the magnetic field geometry, the location of the dominant emission region, and the effects of Faraday rotation and Faraday depolarization along the line of sight. By analyzing how polarization signatures respond to different electron-heating prescriptions and the existence of radiative cooling, we can identify the key physical processes that most strongly influence the resulting polarization maps. This approach enables us to place meaningful constraints on the microphysical mechanisms governing electron heating and transport in radiatively inefficient accretion flows, and to connect polarization observables to the underlying plasma physics encoded in GRMHD simulations.

We organized the paper as follows. We introduce the numerical setup of GRMHD simulations and polarised GRRT in \S \ref{sec:method}. We present our polarization images, as well as the $\beta_2$ shift in the presence of Faraday rotation and higher-order photons for the non-cooling and cooling models in \S~\ref{sec:result}. We further present the polarization asymmetry, variability and radius-dependent linear polarization patterns in \S~\ref{sec:analysis}. In \S~\ref{sec:discussion}, we discuss our findings regarding the physics of electron heating and the implications on observations. Lastly, a conclusion is given in \S~\ref{sec:conclusion}.

\section{Model setup and parameters}
\label{sec:method}

\subsection{General relativistic magnetohydrodynamic simulations}

\begin{table*}
\centering
\begin{tabular}{ccccccccccc}
\hline
$a_*$ & ID & Heating & $i$ & $\rho_{\rm unit}$/$10^{20}$ & $\dot{m}$ & $t_s$ & $t_f$ & $\Delta t$ & Resolution & Cooling \\
 &  &  &   (deg) &  &  ($\dot{M}_{\rm Edd}$) & (M) & (M) & (M) &  &  \\
\hline
\multicolumn{11}{c}{Non-cooling Models} \\
\hline
0.9375 & NC\_MR &  Reconnection & 163 & 3.29 & -- & 10000 & 15000 & 10 & $384 \times 192 \times 192$ & no \\
0.9375 & NC\_MR &  Reconnection & 163 & 16.45 & -- & 10000 & 15000 & 10 & $384 \times 192 \times 192$ & no \\
0.9375 & NC\_MR &  Reconnection & 163 & 32.91 & -- & 10000 & 15000 & 10 & $384 \times 192 \times 192$ & no \\
0.9375 & NC\_KA &  Turbulence & 163 & 3.29 & -- & 10000 & 15000 & 10 & $384 \times 192 \times 192$ & no \\
0.9375 & NC\_KA &  Turbulence & 163 & 16.45 & -- & 10000 & 15000 & 10 & $384 \times 192 \times 192$ & no \\
0.9375 & NC\_KA &  Turbulence & 163 & 32.91 & -- & 10000 & 15000 & 10 & $384 \times 192 \times 192$ & no \\
\hline
\multicolumn{11}{c}{Cooling Models} \\
\hline
0.9375 & C\_MR-6 & Reconnection  & 163 & 3.29 & $1.0 \times 10^{-6}$ & 10000 & 15000 & 10 & $384 \times 192 \times 192$ & yes \\
0.9375 & C\_MR5e6 & Reconnection  & 163 & 16.45 & $5.0 \times 10^{-6}$ & 10000 & 15000 & 10 & $384 \times 192 \times 192$ & yes \\
0.9375 & C\_MR-5 & Reconnection  & 163 & 32.91 & $1.0 \times 10^{-5}$ & 10000 & 15000 & 10 & $384 \times 192 \times 192$ & yes \\
0.9375 & C\_KA-6 & Turbulence  & 163 & 3.29 & $1.0 \times 10^{-6}$ & 10000 & 15000 & 10 & $384 \times 192 \times 192$ & yes \\
0.9375 & C\_KA5e6 & Turbulence  & 163 & 16.45 & $5.0 \times 10^{-6}$ & 10000 & 15000 & 10 & $384 \times 192 \times 192$ & yes \\
0.9375 & C\_KA-5 & Turbulence  & 163 & 32.91 & $1.0 \times 10^{-5}$ & 10000 & 15000 & 10 & $384 \times 192 \times 192$ & yes \\
\hline
\end{tabular}
\caption{Overview of GRMHD and radiative transfer models employed to produce various time series of synthetic images. \label{table:sum}}
\end{table*}

We used a set of three-dimensional (3D) GRMHD simulations of a magnetized torus around a spinning black hole using the {\tt BHAC} code \citep{Porth2017ComAC, Olivares2019A&A}, adopting the two-temperature framework of \cite{Dihingia2023MNRAS} with both Coulomb coupling \citep{Spitzer1965pfig.book, Colpi1984ApJ} and radiative cooling \citep{Esin1996ApJ, Narayan1995ApJ} are also included. 
The electron temperature is obtained independently by solving the electron entropy equation \citep{Ressler2015MNRAS, Mizuno2021MNRAS, Dihingia2023MNRAS} and considering the grid-scale dissipation models to mimic the physics of electron thermodynamics, e.g., turbulence \citep{Kawazura2019PNAS} and magnetic reconnection \citep{Rowan2017ApJ}. 

Each simulation initialize from a Fishbone–Moncrief hydrodynamic equilibrium torus \citep{Fishbone1976ApJ, Uniyal:2024sdv} with parameters $r_{\rm in} = 20 \  R_{\rm g}$ and pressure maximum location $r_{\rm max} = 40 \  R_{\rm g}$, where $R_{\rm g} \equiv GM/c^2 $  is the gravitational radius of the black hole and $M$ its mass.
We use an ideal-gas equation of state for both protons and electrons, characterized by a constant relativistic adiabatic index $\Gamma_{\rm g} =4/3$ \citep{Rezzolla2013rehy}. Weak single-loop magnetic fields are superimposed on this equilibrium torus, with its radial profile chosen to supply sufficient magnetic flux to drive the system into a magnetically arrested disc (MAD) state \citep{Narayan2003PASJ, Tchekhovskoy2011MNRAS}. To seed the magnetorotational instability (MRI), we impose a $4\%$ random perturbation on the gas pressure within the torus. In this paper, we choose a dimensionless spin parameter $a = 0.9375$  for all simulations.

The simulations are performed in spherical Modified Kerr–Schild coordinates, with the outer radial boundary located at $r = 2500\,M$ and the inner boundary of the simulation domain is well inside the black hole horizon in all cases. The effective grid resolution is $384 \times 192 \times 192$, with 3 layers of static mesh refinement, where the highest
resolution is concentrated in $
\lvert \theta - \tfrac{\pi}{2} \rvert \le \tfrac{\pi}{4}$ and $r < 100\,M
$. Initially, the simulations are evolved without Coulomb interactions and radiative cooling until $t = 10000\,M$, by which time the system nearly reaches a quasi-stationary MHD state. Radiative cooling and Coulomb interactions are then switched on with a constant cooling rate. In this work, we explore three values of the dimensionless accretion rates $\dot{m}$, which are normalised to the Eddington rate, $1 \times 10^{-5}$, $5 \times 10^{-6}$, and $1 \times 10^{-6}$. Finally, the simulations are evolved up to $t = 15 000\,M$.

\subsection{Polarised general relativistic radiative transfer}

In this work, we produce the polarization signatures from the GRMHD simulations via the polarised general relativistic radiative transfer code \texttt{BHOSS} \citep{Younsi2016PhRvD, Younsi2020IAUS}.
The \texttt{BHOSS} code is developed from the framework of general relativistic radiative transfer and ray-tracing formalism developed by \citep[][]{Fuerst2004A&A,Younsi2012A&A}. 
It accounts for the evolution of the Stokes parameters through solving the polarised radiative transfer equation along the geodesics of the GRMHD data.
This produces polarimetric images and quantities at given frequencies, allowing us to compare with polarimetry observations, as those of horizon-scale imaging of the M87 \citep{EHT2021ApJ}.

In the calculations, the transfer coefficients are defined locally. They account for synchrotron emission, absorption \citep[see e.g.][]{Pandya2016ApJ}, and Faraday rotation \citep[see e.g.][]{Dexter2016MNRAS}. 
The coefficients, which are specified by a local frequency for a pitch angle with respect to the local magnetic field, depend on the magnetic field and the properties of the charged particles, in particular, the electron temperature(s)  and density (for detailed discussion of the radiative transfer coefficients in polarised radiative transfer calculations in 4-Stokes, see \citet{Pacholczyk1977Book,Jones1977ApJa,Jones1977ApJb}). 
All the quantities essential for the polarised radiative transfer calculations are derived/imported from the GRMHD data.

In the two-temperature GRMHD simulations, the electron temperature is directly from the electron-heating prescriptions. 
The dimensionless electron temperature is in the form of 
\begin{equation*}
    \Theta_\mathrm{e} = \left( \frac{p_\mathrm{e}}{\rho} \right) \left( \frac{m_\mathrm{p}}{m_\mathrm{e}} \right) \ ,
\end{equation*}
where $p_\mathrm{e}$, $\rho$, $m_\mathrm{p}$, and $m_\mathrm{e}$ are the electron pressure, fluid rest-mass density, proton mass, and electron mass. The electron temperature in $c.g.s$ units is expressed as $T_\mathrm{e} = m_\mathrm{e} c^2 \Theta_\mathrm{e}/k_\mathrm{B}$, where $k_\mathrm{B}$ is the Boltzmann constant. 

For our purpose of study, we apply the simplest electron density function (eDF), the thermal distribution. The eDF might not affect the polarization signatures as much as the intensity spectrum in 230~GHz or 86~GHz. 
In the optically thin region roughly between $10^9$ and $10^{11}$ Hz, the intrinsic polarization of synchrotron emissions is around $70 \%$, regardless of the electron density functions \citep{Paul2025MNRAS}. Certainly, for a more complicated environment performed by GRMHD simulations, non-uniform local magnetic fields and anisotropic electron momentum distributions can deviate from the $70 \%$ polarization, thus further investigation is still needed. We thus apply a hybrid kappa distribution with thermal and non-thermal contributions \citep{Xiao2006PPCF, Pandya2016ApJ}
\begin{align}
\frac{{\rm d} n_{\mathrm{e}}}{{\rm d} \gamma_{\mathrm{e}}}= N \gamma_{\mathrm{e}} \;\! 
\sqrt{\gamma_{\mathrm{e}}^2-1}
 \;\! \left(1+\frac{\gamma_{\mathrm{e}}-1}{\kappa w}\right)^{-(\kappa+1)} \ , 
\label{eq:dn-dgamma}
\end{align}
where $N$ is the normalization factor. For relativistic electrons, the distribution can be simplified as $d n_\mathrm{e}/d \gamma_\mathrm{e} \propto \gamma_\mathrm{e}^{-(\kappa -1 )}$. In this study, we apply a constant kappa value of $3.5$. Although the low value of $\kappa$ represents an overestimation of the non-thermal distribution, it serves as a proof of concept.

We took M87 as our study target. This means a set of input physical parameters for GRRT calculations, including the inclination angle of $163^{\circ}$ and black hole mass of $M_\mathrm{BH} = 6.5 \times 10^9 M_{\odot}$ at the distance of
16.8~Mpc \citep{EHT2019ApJ...875L...5E}. However, we did not scale the density unit because we set the mass accretion rate for the calculation of radiative cooling so that the total intensity does not satisfy the observed range of $0.5$ Jy. 
Instead, to study the electron effect coherently with radiative cooling and Comloub coupling, we apply a specific scaling factor corresponding to set mass accretion rates $\dot{m}$ from GRMHD simulations, while for the purpose of comparison, we use the same scaling density factor for GRMHD simulations without radiative cooling. Other parameters related to GRMHD simulations are the $\sigma_{\rm cut}$, which we choose $\sigma_{\rm cut} = 1$ to avoid a simulation domain with strong magnetization. To study different emission regions, we use the Bernoulli parameter $-h u_t < 1.02$ for disk region and $-h u_t > 1.02$ for the jet-sheath region, respectively. Using the parameters above, we produce polarised snapshots between $t = 12000\,M$ and $t = 15000\,M$ with a $10\,M$ cadence and a resolution of ($512 \times 512$), with a half field of view $114.6\mu as$ ($=30 \ R_{\rm g}$) at 86~GHz and $64.9 \mu as$ ($=17 \ R_{\rm g}$) at  230~GHz respectively. All GRMHD and radiative transfer models are summarized in Table~\ref{table:sum}.

\section{Polarimetric Images and properties}
\label{sec:result}

To characterize the polarization signatures of emission originating near the black hole horizon, we analyze the electric vector position angle (EVPA) maps and polarization observables commonly employed in Event Horizon Telescope (EHT) analyses \citep{EHT2021ApJ2}. In particular, we focus on image-domain EVPA morphology and global polarization diagnostics, including the rotationally symmetric decomposition coefficient $\beta_2$.

\subsection{Snapshot of GRMHD Image and Polarimetric Map}

\begin{figure*}
\centering
\includegraphics[width=2.0\columnwidth]{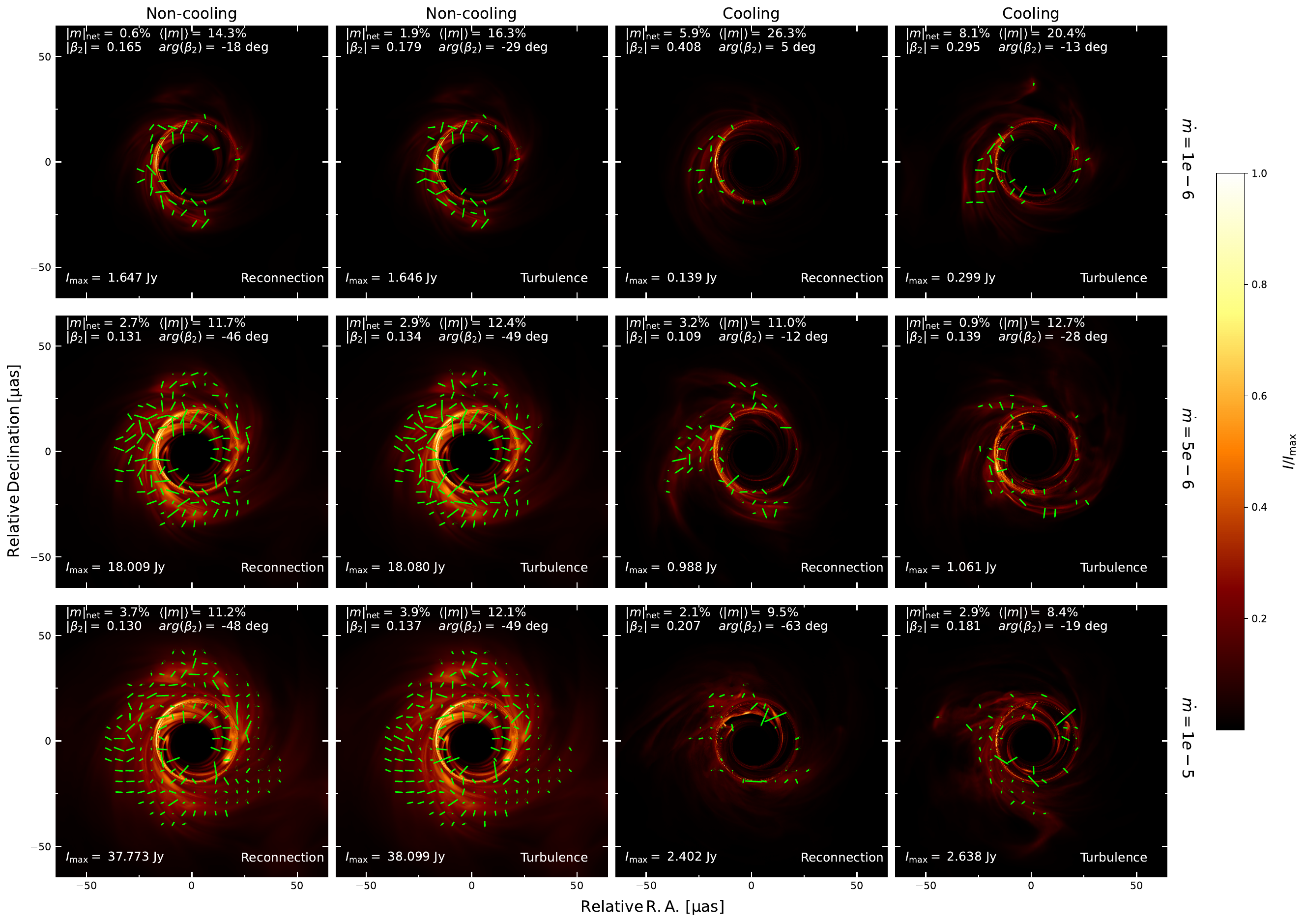}
\caption{Intensity images with EVPA patterns at snapshot $t=14800\,M$ at 230~GHz. At each panel, the GRRT images are produced by the  post-processing of various two-temperature MAD simulations at $a=0.94$ and 
inclination angle of $163^{\circ}$, with Turbulence  and Reconnection heating, with or without cooling (electron feedback), and at accretion rates $\dot{m}$ of $1 \times 10^{-6}$, $5 \times 10^{-6}$, and $1 \times 10^{-5}$ (from top to bottom). Polarization ticks only displayed in regions where $I>0.1\,I_\mathrm{max}$.  The black hole spinning axis rotates $108^{\circ}$ counter-clockwise to match the observed appearance of  M87. The eDF used here are thermal distributions, and all images are produced from the total region of GRMHD simulations.}
\label{fig:I}
\end{figure*}

Figure~\ref{fig:I} presents the intensity images and linear polarization maps for the MAD simulations at a single snapshot, spanning different electron heating prescriptions (turbulent and reconnection heating), the inclusion or exclusion of electron feedback (here we call the feedback as cooling), and a range of mass accretion rates. In all panels, the intensity is normalized by the maximum image intensity, whose value is reported in the lower-left corner of each subplot. Linear polarization is visualized using EVPA ticks overlaid on the intensity maps; the tick orientation corresponds to the electric vector position angle, and the tick length scales with the polarised intensity $P=\sqrt{Q^2+U^2}$. For each image, we report the image-integrated polarization fraction $|m_{\rm net}|$, the image-averaged polarization fraction $\langle |m| \rangle$, and the rotationally symmetric decomposition coefficient $\beta_2$. We also show the images convolved with a $20\,\mu\mathrm{as}$ circular Gaussian beam in Fig.~\ref{fig:blur}.

In this snapshot, we find that the differences between electron heating prescriptions are negligible in the non-cooling simulations. At a fixed accretion rate, models employing reconnection heating yield a similar brightness as shown by the peak intensities and more spatially extended EVPA structures, while the overall polarization fractions remain similar between the two heating prescriptions, with slightly lower polarization compared to the models using turbulence heating. 
When electron cooling and Coulomb coupling are included, the total emission is significantly reduced, and the intensity and polarization morphologies become broadly differentiable across heating models at the same accretion rate. With increasing mass accretion rates, the emission inside the photon ring region becomes brighter. This is consistent with the GRMHD picture, where the cooling is much more efficient in the disk region than in the jet or jet-sheath regions.

$\angle \beta_2$ of all the panels has a value closer to 0 with a mild tendency to negativity. This is consistent with a general left-handed radial EVPA pattern, which is typically seen in a MAD simulation at high spin $a=0.9375$, with a right-handed plasma rotation direction. 
Non-cooling systems show a consistent decrease of $|\beta_2|$ with increasing $\dot{m}$, and a more randomized EVPA pattern. The randomness is most visible at the near-horizon disk region, while the EVPA pattern shows a rotationally symmetric spiral feature at the outer disk region. Together, the EVPA patterns and $\beta_2$ indicate the effect of magnetic field structure. Other polarization properties are summarised in Fig.~\ref{fig:Hist_pol}.

We briefly note that non-thermal electron distributions are often invoked to explain enhanced jet emission. In our study, even adopting a relatively strong non-thermal component ($\kappa=3.5$), we find that polarization properties at $230\,\mathrm{GHz}$ are largely insensitive to the choice between thermal and hybrid kappa electron distributions. Differences become more pronounced at $86\,\mathrm{GHz}$, but since this work focuses on $230\,\mathrm{GHz}$, we use thermal electron distributions throughout. A detailed comparison is presented in Figure~\ref{fig:Peak_I} and discussed further in Appendix~\ref{app:distribution}.

\subsection{Images without Faraday Rotation}

\begin{figure}
\centering
\includegraphics[width=1.01\columnwidth]{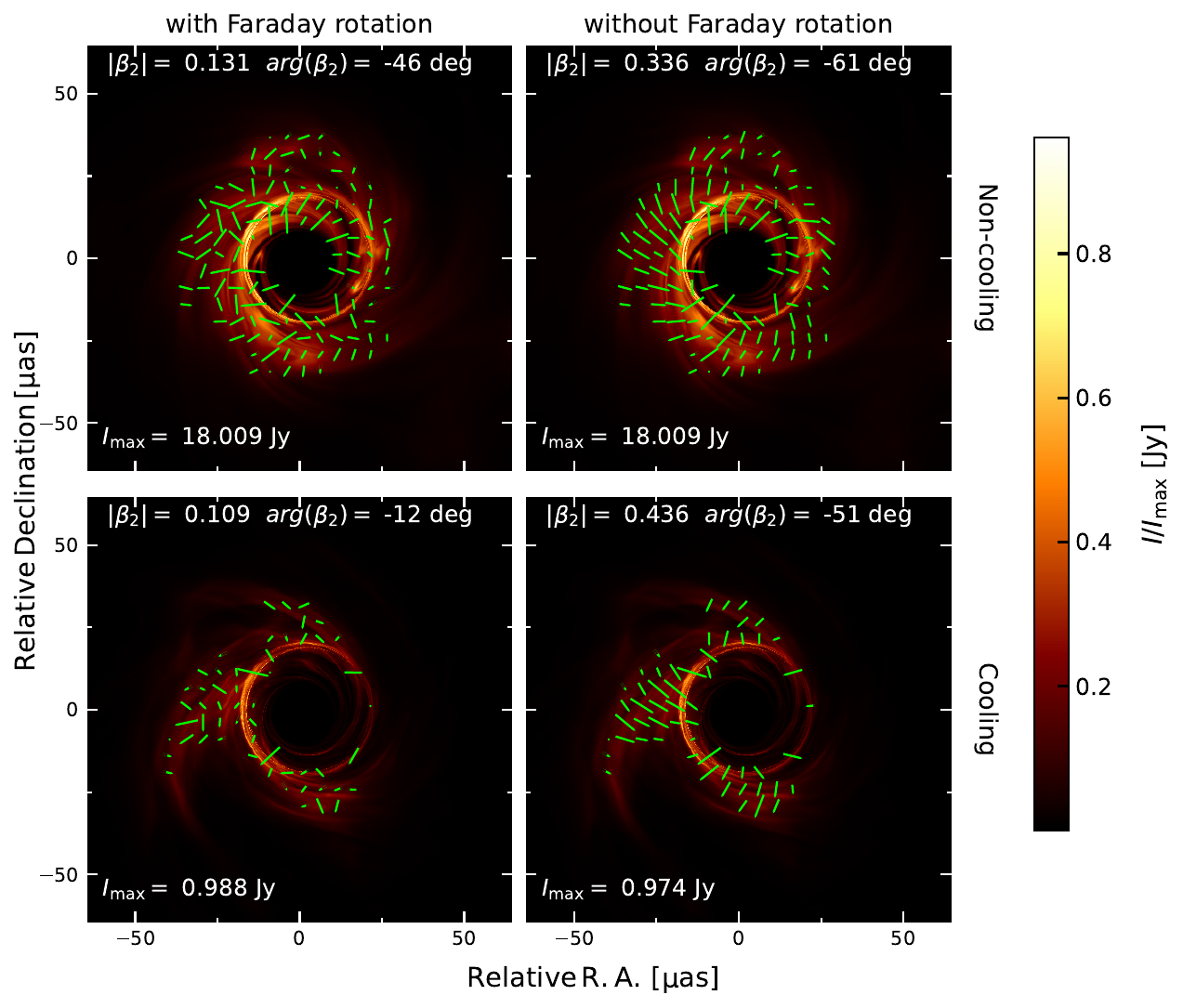}
\caption{Comparisons of images between original images with Faraday rotation and images excluding Faraday rotation. This illustration is for systems at $\dot{m}=5 \times 10^{-6}$ with a reconnection heating prescription.}
\label{fig:faraday}
\end{figure}

Figure~\ref{fig:faraday} compares EVPA maps computed with and without Faraday rotation for a representative system with $\dot{m}=5\times10^{-6}$ using the reconnection heating prescription. By excluding Faraday rotation, these images isolate the intrinsic polarization structure imprinted at the emission region by the magnetic field geometry and electron thermodynamics, prior to any propagation effects due to Faraday rotation along the line of sight.

Removing Faraday rotation reveals substantially more coherent and ordered EVPA structures, particularly in the disk-dominated regions. The differences are most pronounced in the inner disk, where plasma density and magnetic field strength are highest. The images that included Faraday rotation exhibit systematic EVPA rotations on small spatial scales together with shortened polarization vectors, indicative of strong differential Faraday rotation and associated depolarization. When Faraday rotation is excluded, these effects disappear, and the EVPA ticks align smoothly, tracing the underlying magnetic field configuration more clearly. 
This yields a stronger image-integrated polarization, as indicated by a larger value of $|\beta_2|$, and a more coherent toroidal rotation, reflected in a more negative $\angle \beta_2$.

In both non-cooling and cooling cases, the region interior to the photon ring shows little sensitivity to the inclusion or exclusion of Faraday rotation. This suggests that the polarization signal originating from these regions is either produced in relatively Faraday-thin plasma or dominated by geometric and relativistic effects that are largely unaffected by propagation through the surrounding medium. We will investigate the spatial difference of the EVPA patterns in the study of radial-dependent $\beta_2$ in Sec.~\ref{sec:radius-beta2}.

\begin{figure}
\centering
\includegraphics[width=1.\columnwidth]{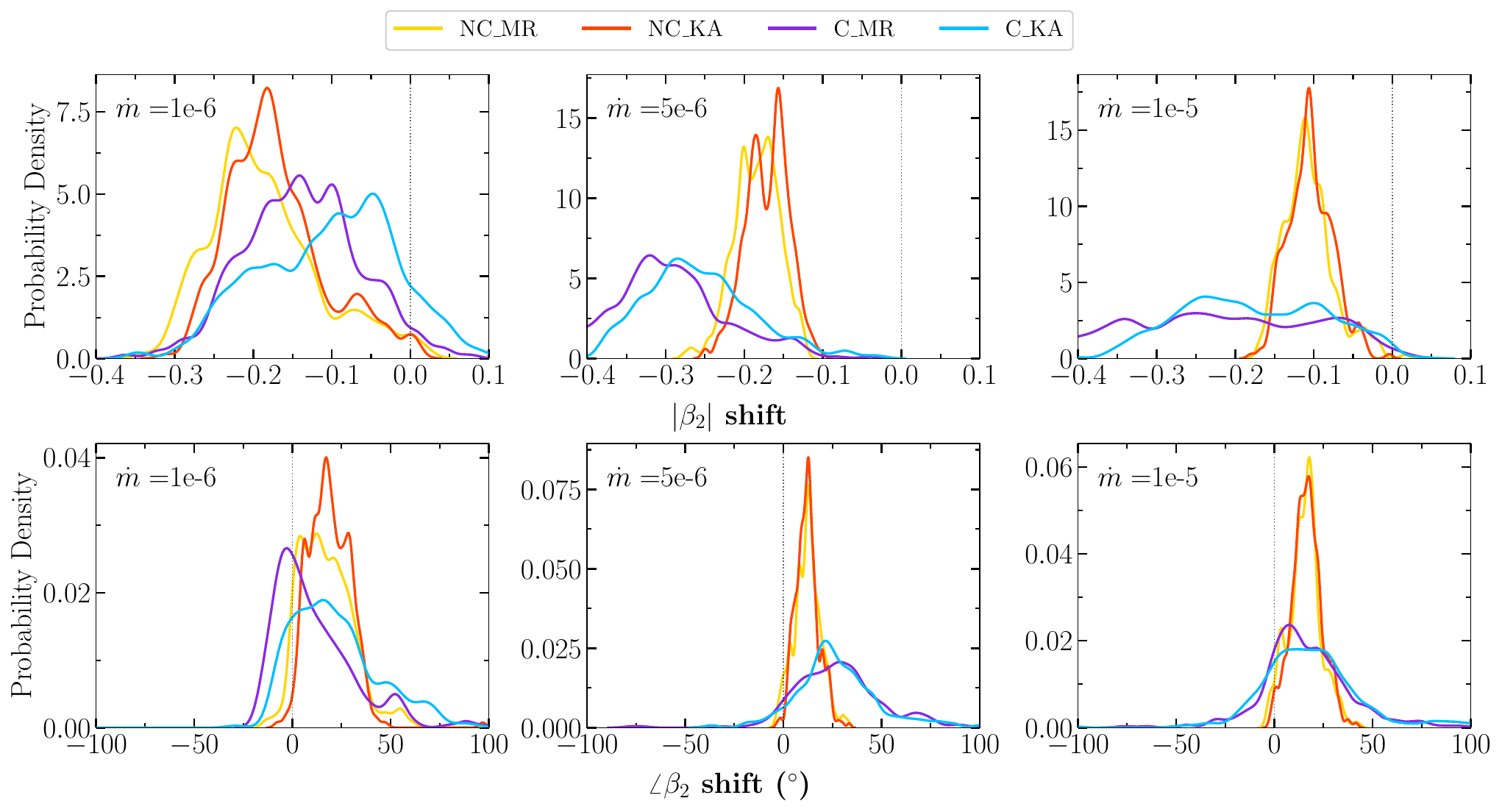}
\caption{$\beta_2$ phase shift due to Faraday rotation. Original minus images without Faraday rotation. The top and bottom panel shows the shift of $|\beta_2|$ and $\angle \beta_2$. From left to right, each panel presents the systems at accretion rates $\dot{m}$ of $1 \times 10^{-6}$, $5 \times 10^{-6}$, and $1 \times 10^{-5}$, respectively. Different GRMHD with different electron treatments are colored as gold, red, purple, and blue, for non-cooling reconnection, non-cooling turbulence, cooling reconnection, and cooling turbulence, respectively.  Excluding Faraday rotation reveals the intrinsic difference of polarization emissions from cooling cases and from non-cooling cases.}
\label{fig:dbeta_Faraday}
\end{figure}

To demonstrate that these trends are robust and not coincidental features of a single snapshot, we quantify the effect of Faraday rotation using the second azimuthal polarization mode, $\beta_2$, in Figure~\ref{fig:dbeta_Faraday}. We compute the shifts in $|\beta_2|$ and $\angle\beta_2$ as the difference between values obtained from original images with Faraday rotation and those computed from images without Faraday rotation, evaluated over the full time series of all GRMHD simulations considered. The resulting distributions are shown as probability density functions,  where the area under each curve is normalized to unity and the distribution width reflects the distributions of the Faraday-induced shifts over the simulation period. The zero shift indicated by the light grey vertical line. 

Faraday rotation leads to a systematic reduction in $|\beta_2|$, demonstrating that Faraday depolarization efficiently suppresses the coherent quadrupolar polarization mode, and thus the image-integrated polarization magnitudes decrease. At the lowest mass accretion rate $\dot{m}=10^{-6}$, the shift of $|\beta_2|$ due to the effect of Faraday rotation is comparable between the non-cooling and cooling models. However, at higher mass accretion rates $\dot{m}=5 \times 10^{-6}$ and $\dot{m}=10^{-5}$, the effect of Faraday rotation becomes weaker for non-cooling cases, whereas the cooling cases become stronger.  At the same time,  $\angle\beta_2$ shifts toward larger absolute values, with images excluding Faraday rotation exhibiting more negative values of $\angle\beta_2$.  This behaviour reflects a more coherent toroidal polarization structure that emerges when Faraday scrambling along the line of sight is absent.

 The cooling models also exhibit a broader distribution and larger magnitude of the shifts in $\beta_2$, particularly at higher mass accretion rates. This trend suggests that electron cooling produces more heterogeneous plasma and magnetic-field structures, increasing the susceptibility of the polarised emission to Faraday rotation and depolarization effects.


\subsection{Order and domain decomposed image}
\label{sec:decompose}

GRMHD simulations are a complicated system across multi-scale physics. Synchrotron photons propagate through extended regions of the accretion flow, accumulating geometric, relativistic, and plasma propagation effects along their trajectories before reaching the observer. As a result, polarization images constructed from the full radiative transfer encode a superposition of signals originating from distinct photon paths and physical domains, which obscures the origin of specific polarization features. To recover this information, we adopt a physics-motivated decomposition of the images by photon order and emission domain.

We first separate direct (0th-order) images from higher-order images ($n>0$). Higher-order photons orbit the black hole one or more times before escaping and therefore experience stronger gravitational lensing, frame dragging, and path-dependent polarization transport, mixing spacetime and plasma effects. In contrast, 0th-order photons are less affected by strong-field lensing and more directly trace the local magnetic field geometry and plasma motion at the emission site. After isolating the 0th-order images, we further decompose the emission by physical domain using the Bernoulli parameter $-h u_t$. We identify gravitationally bound plasma ($-h u_t < 1.02$) as disk material and unbound plasma ($-h u_t > 1.02$) as jet and jet-sheath regions. While this criterion is necessarily approximate, it provides a physically motivated separation between disk- and jet-dominated emission.

\begin{figure*}
\centering
\includegraphics[width=2.2\columnwidth]{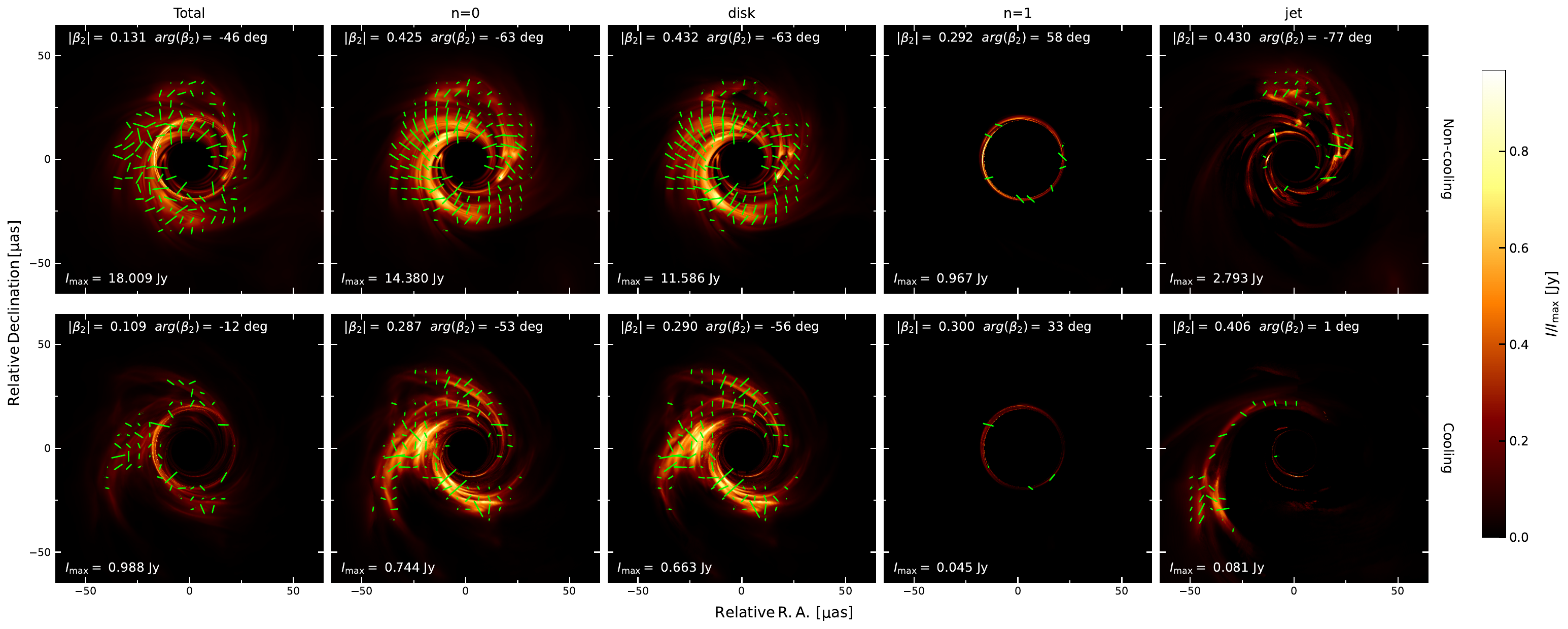}
\caption{Image decompositions using the same representative system shown in Figure~\ref{fig:faraday}, with the non-cooling model in the top row and the cooling model in the bottom row. From left to right, we show the total image, the 0th-order image, the 1st-order image, and the 0th-order disk- and jet-domain images.}
\label{fig:decompose}
\end{figure*}

\begin{figure}
\centering
\includegraphics[width=1.0\columnwidth]{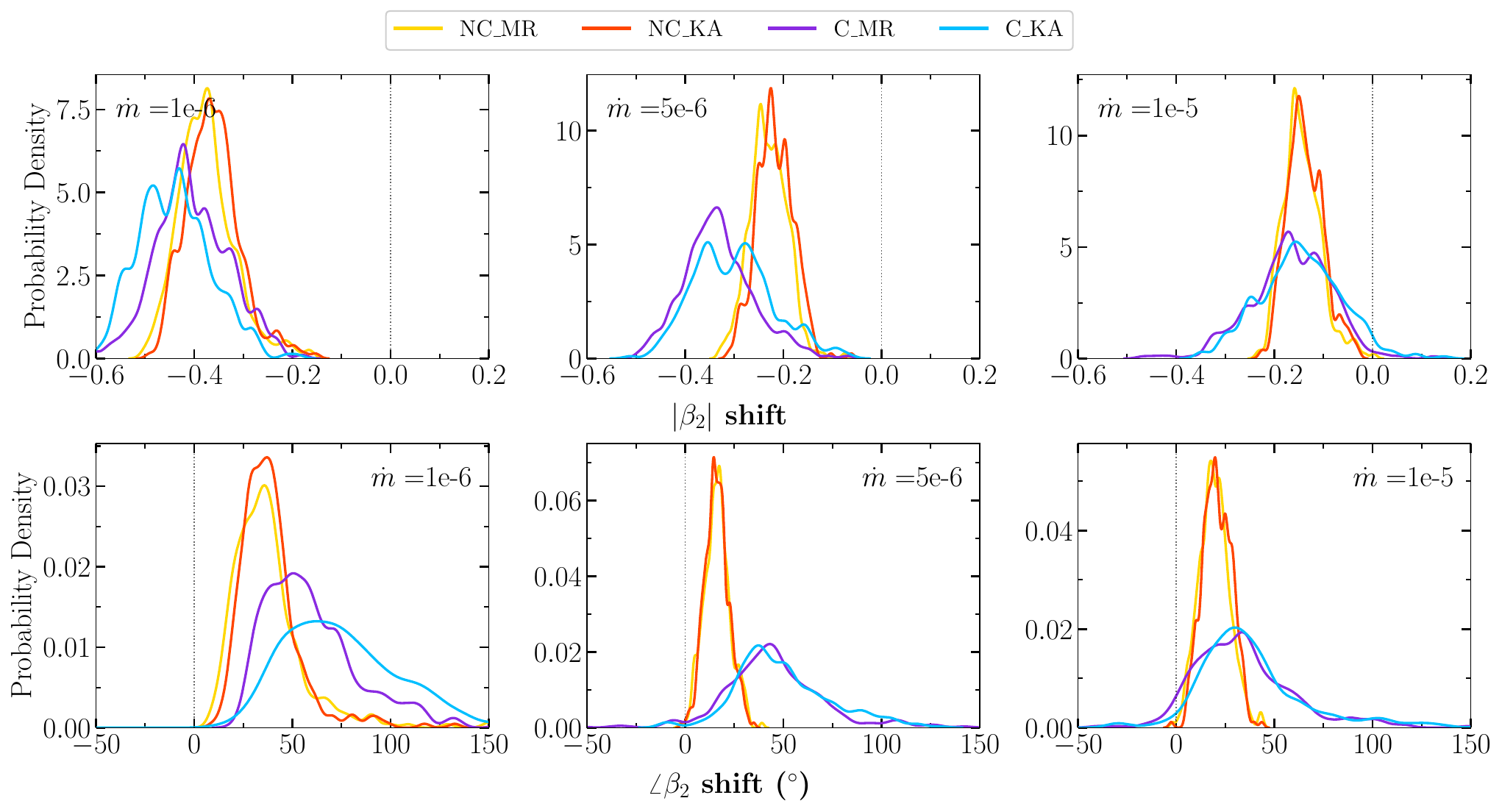}
\caption{Same as Fig.~\ref{fig:dbeta_Faraday} but $\beta_2$ phase shift due to higher-order emissions. Original minus images without higher-order photons.}
\label{fig:dbeta_n=0}
\end{figure}

Figure~\ref{fig:decompose} illustrates the order- and domain-separated images for the same representative system shown in Figure~\ref{fig:faraday}, with the non-cooling model in the top row and the cooling model in the bottom row. From left to right, we show the total image, the 0th-order image, the 1st-order (photon-ring) image, and the disk- and jet-domain images. The 0th-order images exhibit more coherent EVPA structures than the total images and closely resemble the images computed without Faraday rotation in Figure~\ref{fig:faraday}. In both cases—removing  higher-order photons or removing Faraday rotation—the resulting EVPA patterns are smoother and more globally ordered than in the full images. 
Despite this visual similarity, the underlying physical mechanisms are fundamentally different. Faraday rotation is a plasma propagation effect whose strength depends sensitively on electron density and temperature, whereas higher-order emission reflects gravitational lensing and polarization transport in the strong-field regime. This distinction becomes clear when comparing the quantitative behaviour of the polarization modes. 
As shown in Figure~\ref{fig:dbeta_n=0}, the shifts in $\angle\beta_2$ induced by removing higher-order photons behave broadly similarly across mass accretion rates, in contrast to the strong $\dot{m}$ dependence seen for Faraday rotation in Figure~\ref{fig:dbeta_Faraday}. This indicates that polarization mixing due to spacetime effects is largely insensitive to plasma density, unlike Faraday rotation.

At low mass accretion rates, cooling models exhibit a stronger reduction in $|\beta_2|$ when higher-order photons are removed than non-cooling models. This behaviour is directly related to the relative importance of photon-ring emission. As shown in the third column of Figure~\ref{fig:decompose}, the disk emission is dimmer in the cooling case, increasing the fractional contribution of the photon ring to the total polarised flux. In this representative example, the photon-ring contribution reaches $4.3\%$ in the cooling model, compared to $3.3\%$ in the non-cooling model. As the disk density and brightness increase with $\dot{m}$, the relative contribution of the photon ring decreases, and the shifts in $|\beta_2|$ for cooling and non-cooling models converge. This behaviour indicates that, once the photon-ring contribution is sufficiently diluted by disk emission, the gravitational effect on polarization becomes effectively independent of the plasma thermodynamics.

\begin{figure}
\centering
\includegraphics[width=1.0\columnwidth]{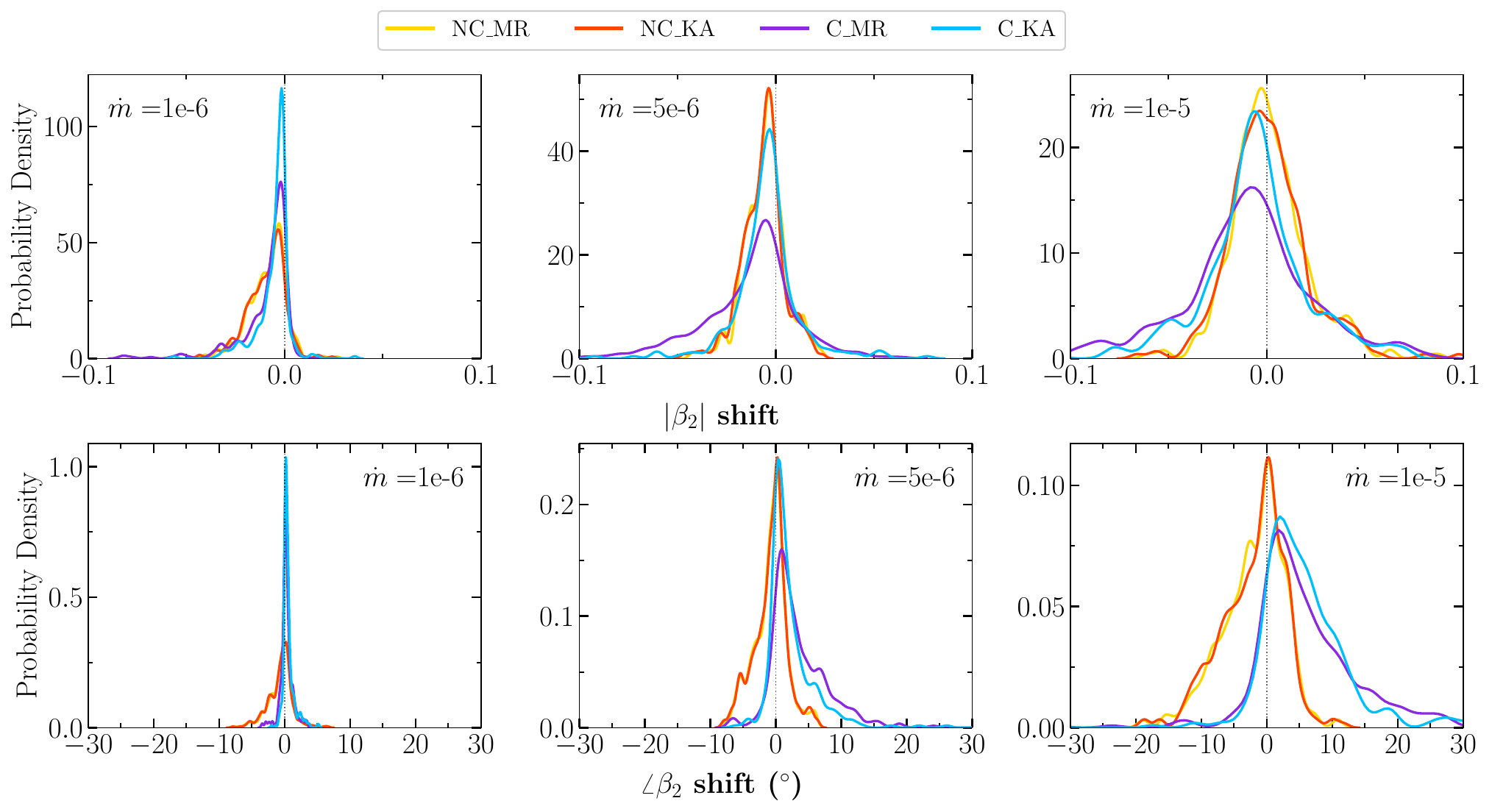}
\caption{Same as Fig.~\ref{fig:dbeta_Faraday} but $\beta_2$ phase shift due to emissions from jet region. $n=0$ order images minus 0th-order disk images.}
\label{fig:dbeta_disk}
\end{figure}

The domain decomposition further clarifies the origin of residual differences. The disk-only 0th-order images are visually indistinguishable from the total 0th-order images and exhibit nearly identical $\beta_2$ values, demonstrating that direct emission is overwhelmingly disk-dominated. The primary difference is the removal of a localized hotspot in the upper-left quadrant, which originates in the jet-sheath region and displays a spiral emission structure. This feature contributes a small additional polarised component and likely explains the modest shift in $|\beta_2|$ observed at higher mass accretion rates in Figure~\ref{fig:dbeta_disk}. However, because the jet-sheath polarization follows the same frame-dragging–induced EVPA rotation as the disk emission, the phase $\angle\beta_2$ remains largely unchanged, especially for non-cooling cases.

\section{Analysis of linear polarization patterns}
\label{sec:analysis}

We focus on  $\beta_2$ to probe the plasma properties in GRMHD simulations that incorporate various parts of electron thermodynamics. $\beta_2$ did a great work for revealing the rotationally symmetric linear polarization pattern, as in the case of polarimetric observations of M87 and Sgr~A*. The second order Fourier mode coefficient $\beta_2$ is computed as \citep{Palumbo2020ApJ}:
\begin{align}
\beta_m & 
=\frac{1}{I_{\operatorname{ann}}}
\int_{\rho_{\min }}^{\rho_{\max }}
 \mathrm{d} \rho  \; \rho 
\int_0^{2 \pi} \mathrm{~d} \varphi \  P(\rho, \varphi) e^{-i m \varphi} \\
I_{\operatorname{ann}} & 
=\int_{\rho_{\min }}^{\rho_{\max }} \mathrm{d} \rho\; \! \rho 
\int_0^{2 \pi} \mathrm{~d} \varphi \
  I(\rho, \varphi) 
\end{align}
where $P(\rho, \varphi)$ and $I(\rho, \varphi)$ are the complex linear polarization and intensity located at the polar image-plane coordinates $(\rho, \varphi)$. The magnitude $|\beta_2|$ quantifies the strength of the rotationally symmetric polarization, while the phase $\angle \beta_2$ specifies the orientation of this symmetry relative to a purely radial EVPA pattern. For example, $\angle \beta_2=0$ corresponds to a purely radial polarization field, $\angle \beta_2 = \pi$ to a purely azimuthal field, and negative or positive $\angle \beta_2$ values indicate spirals with left or right-handed helicity.

\subsection{Decomposition modes of linear polarization patterns}

Although the quadrupolar mode $\beta_2$ is the most commonly used coefficient for characterizing linear polarization patterns—particularly in the context of EHT observations—it is instructive to examine the full set of azimuthal decomposition modes to quantify deviations from rotational symmetry. Higher- and lower-order modes encode asymmetric EVPA structures that are not captured by $\beta_2$ alone and, therefore, can provide additional diagnostic power.

\begin{figure*}
\centering
\includegraphics[width=2.0\columnwidth]{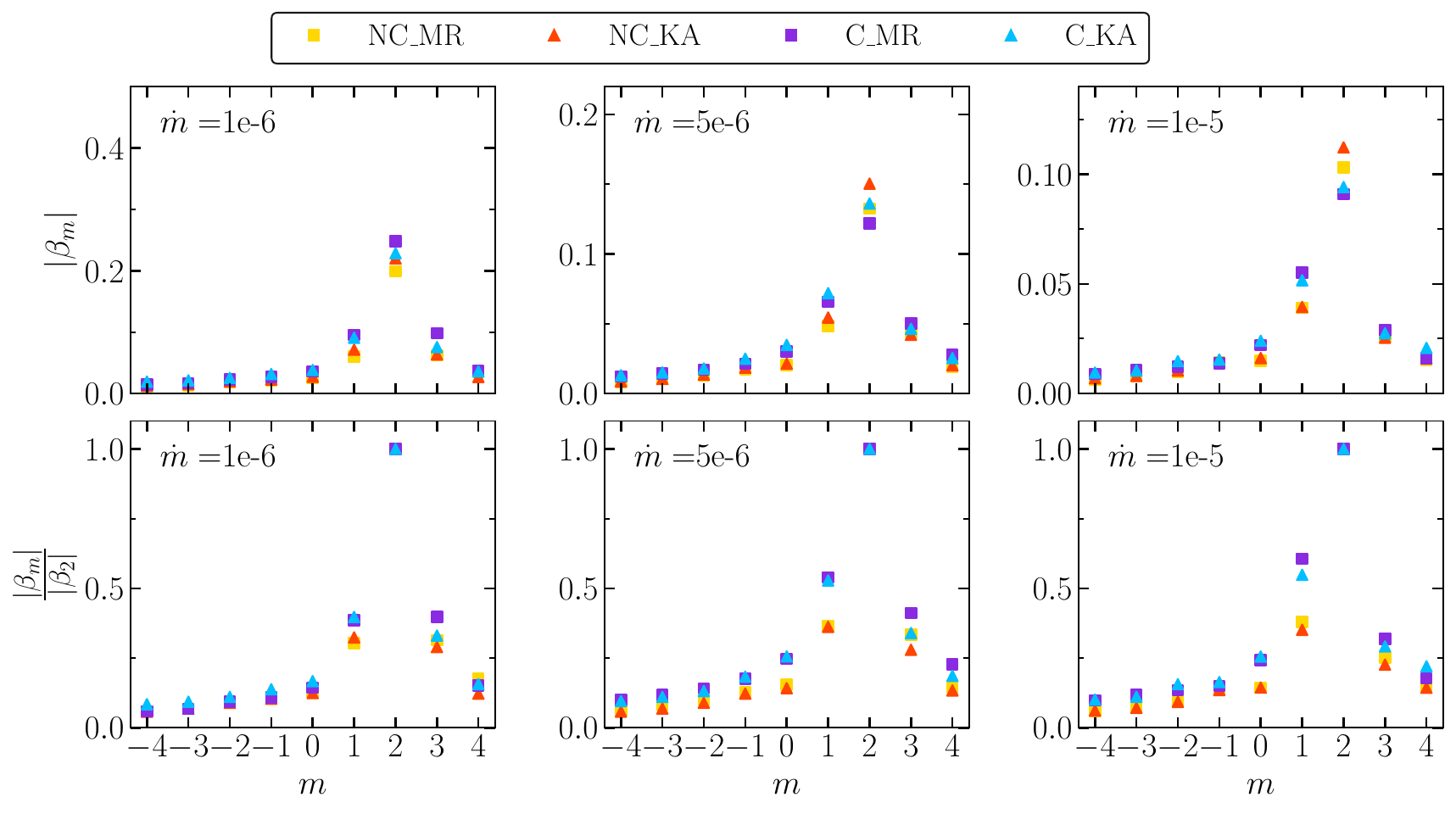}
\caption{Top panel and bottom panel show $\beta_m$ and ratio of $|\beta_m| / |\beta_2|$ at different modes for total emissions.}
\label{fig:beta_m_total}
\end{figure*}

Figure~\ref{fig:beta_m_total} shows the time-averaged decomposition coefficients $\beta_m$ for all simulated models, evaluated over modes $-4 \leq m \leq 4$. The top row presents the absolute amplitudes $|\beta_m|$, while the bottom row shows each mode normalized by the dominant quadrupolar component, $|\beta_m|/|\beta_2|$. Across all accretion rates and electron prescriptions, the $m=2$ mode is clearly dominant, confirming that the linear polarization patterns are primarily quadrupolar in nature and consistent with large-scale, approximately axisymmetric magnetic field structures. Beyond the dominant $\beta_2$ mode, the $m=1$ mode emerges as the second most significant contribution in all models, reaching more than half the amplitude of $\beta_2$ in several cases. The $m=3$ mode is typically the next largest contributor, while the higher-order modes ($|m|\geq4$) are subdominant. This hierarchy indicates that most deviations from perfect rotational symmetry are captured by low-order asymmetric modes, particularly $m=1$. 

The normalized representation over $\beta_2$ mode in the bottom panels of Figure~\ref{fig:beta_m_total} highlights systematic differences between models. The simulations with radiative cooling, both reconnection-heated (C\_MR) and turbulence-heated (C\_KA), exhibit a larger fractional contribution from asymmetric modes ($m\neq2$) compared to their non-cooling counterparts. This trend holds across all three different accretion rates and reflects the increased asymmetry in the EVPA patterns seen in the image-domain polarization maps (Figure~\ref{fig:I}). In cooling models, radiative feedback modifies the spatial distribution of emissivity, leading to less uniform polarization structures and enhanced power in non-axisymmetric modes. 

These results demonstrate that ratios such as $|\beta_1|/|\beta_2|$ and $|\beta_3|/|\beta_2|$ provide robust quantitative measures of asymmetry in linear polarization patterns. These ratios offer a useful complement to $\beta_2$ itself, enabling discrimination between standard GRMHD models and simulations that include additional physics—such as electron feedback—that systematically break rotational symmetry.

\subsection{Variability and $\beta_2$ angle flipping}
\label{sec:beta2_var}

Since the quadrupolar mode $\beta_2$ remains the dominant component of the linear polarization decomposition across all models, we use it as a representative diagnostic of the global polarization structure. In particular, temporal variations in the phase $\angle\beta_2$ provide a compact measure of changes in the large-scale EVPA morphology during the time evolution of the GRMHD simulations.

\begin{figure*}
\centering
\begin{minipage}[t]{0.48\textwidth}
    \centering
    \vspace{0.05cm}
    \includegraphics[width=0.9\linewidth] {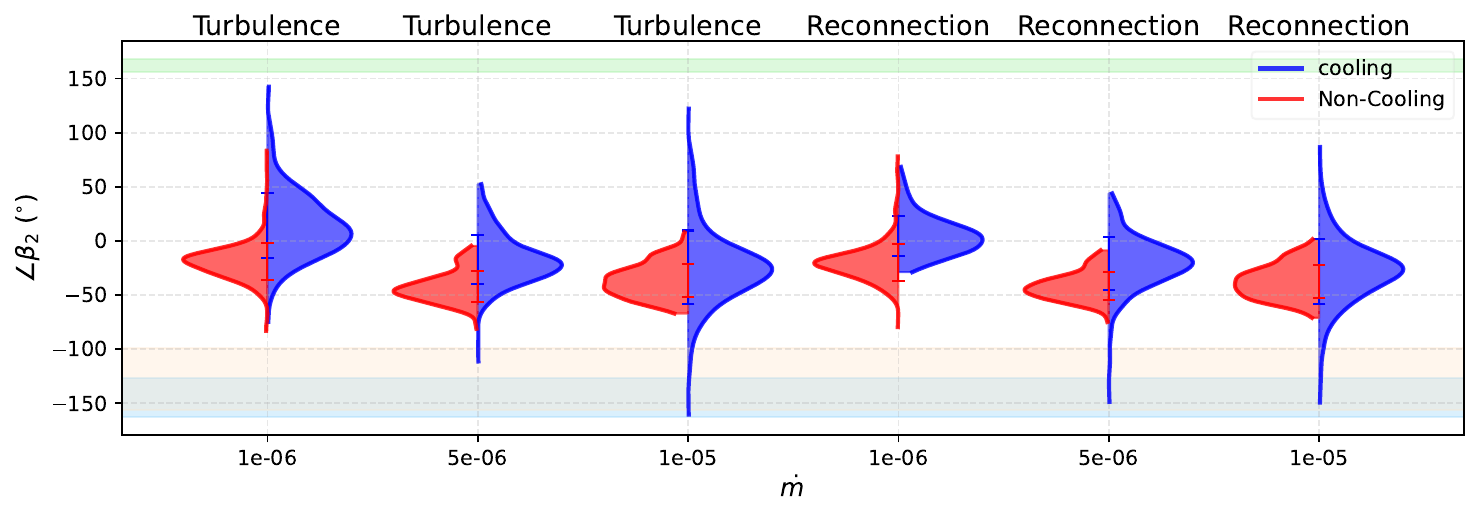}    
    \vspace{0.3cm}
    \hspace{-0.5cm}
    \includegraphics[width=0.95\linewidth]{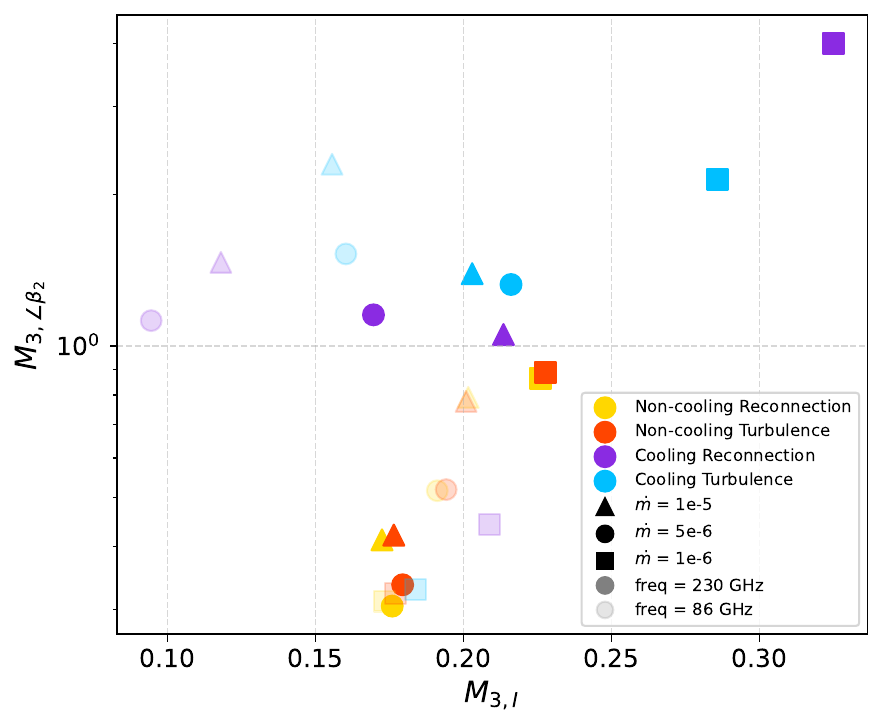}
\end{minipage}
\hfill
\begin{minipage}[t]{0.48\textwidth}
    \centering
    \vspace{0.20cm}
    \hspace{-1.0cm}
    \includegraphics[width=1.1\linewidth]{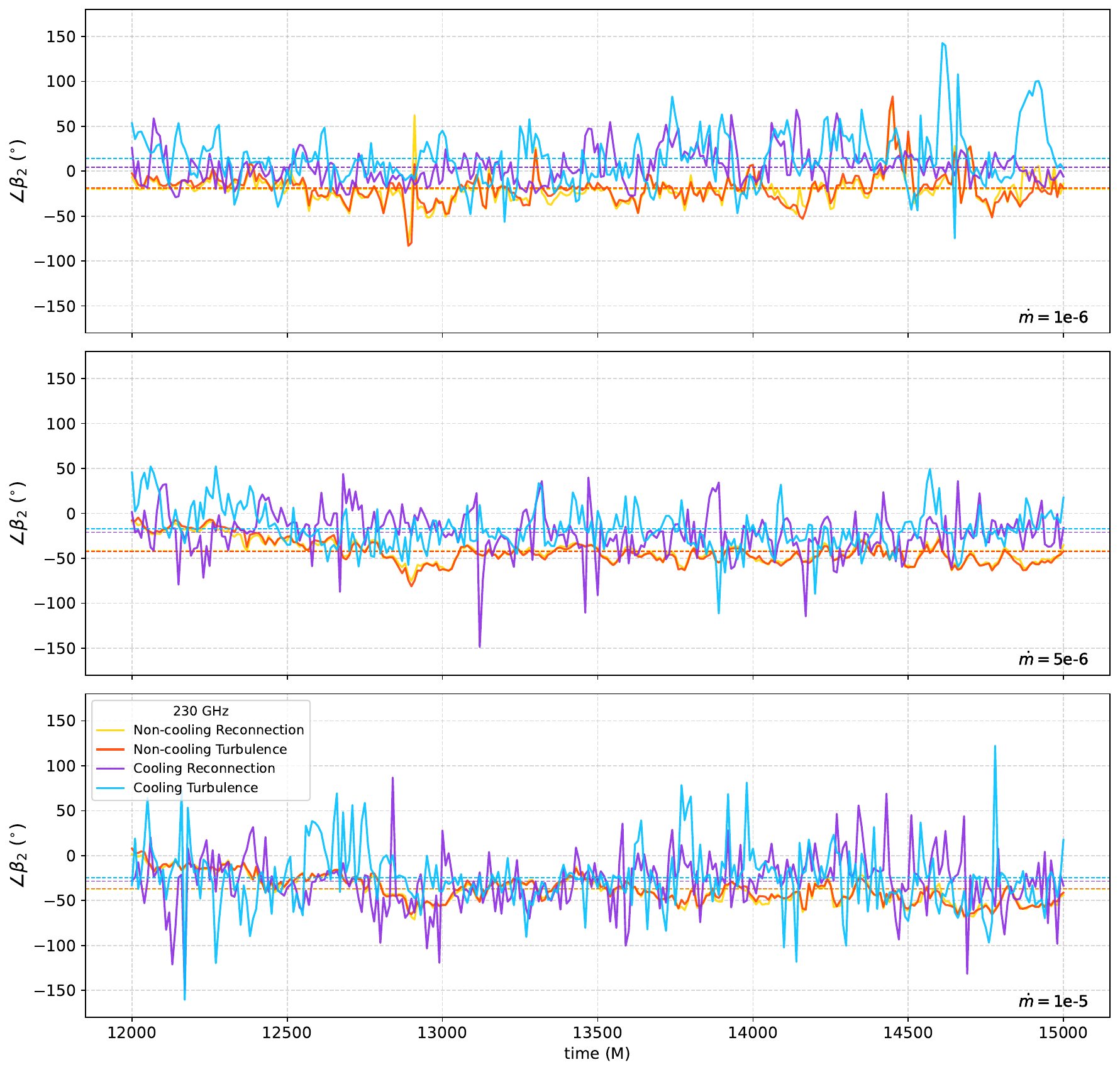}
\end{minipage}
\caption{Top left: Histogram distributions of $\angle\beta_2$. Bottom left: Variability of the rotationally symmetric mode $M_{3,\,\angle\beta_2}$ versus variability of intensity $M_{3,\,I}$, where $M_{3,\,i} = \sigma_i/\mu_i$.
 The standard deviations $\sigma_i$ and mean values $\mu_i$ are calculated from snapshots between $t = 12000\,M$ and $t = 15000\,M$. The radiative properties without cooling are labelled as warm colors of yellow and red, and including cooling with cold colors of purple and blue, with the former using reconnection heating and the latter using turbulence heating. The different mass accretion rates used triangle ($\dot{m} = 1e-5$), circle ($\dot{m} = 5e-6$) and square ($\dot{m} = 1e-6$). The properties at 230~GHz are opaque, while the properties at 86~GHz are transparent. Right: Temporal evolutions of $\angle\beta_2$.}
\label{fig:variability_beta2}
\end{figure*}

Figure~\ref{fig:variability_beta2} summarizes the temporal behaviour of $\angle\beta_2$ using complementary statistical and time-domain diagnostics. The histogram shown in the upper-left panel illustrates the distribution of $\angle\beta_2$ values over the time interval between $t=12000\,M$ and $15000\,M$. Cooling models (shown in blue) exhibit a generally broader distribution, spanning a wide range of angles, whereas non-cooling models (shown in red) remain more tightly clustered around a preferred phase. This indicates that the GRMHD simulations with radiative cooling experience larger amplitude changes in the global EVPA orientation. 

This behaviour is directly visible in the $\angle\beta_2$ light curves shown in the right panels. Cooling models, both turbulence and reconnection cases, display visible temporal excursions and frequent phase reversals (“angle flipping”), most notably at the highest accretion rate $\dot{m}=10^{-5}$. 
We note, however, that phase reversals do not necessarily correspond to the same underlying physical process. The simultaneous evolution of $|\beta_2|$ provides additional information that helps distinguish between different mechanisms responsible for the observed angle flipping.  
For example, asymmetric depolarization, similar to that inferred from the 2017 EHT polarization image of M87*, can suppress the $|\beta_2|$ to very low amplitudes, causing its $\angle\beta_2$ to be highly sensitive to small fluctuations in the polarised signal. In our simulations, the phase reversals such as the ones occurring around $t=12170\,M$ and $t=14780\,M$, are accompanied by exceptionally low values of $|\beta_2|$ that is below 0.05 and should therefore be interpreted with caution. Therefore, the angle flipping may reflect either a genuine change in the global EVPA orientation or a transient loss of polarization coherence highly dependent on $|\beta_2|$ amplitude. The corresponding $|\beta_2|$ light curves are presented in Appendix~\ref{app:beta_2}, where they provide additional diagnostic power for distinguishing between these possibilities.  In contrast, non-cooling models show a comparatively smooth and stable evolution of $\angle\beta_2$, with fluctuations remaining confined within a narrower angular range.

To quantify these trends, we characterize the variability using the dimensionless measure $M_{3,\,i} \equiv \sigma_i / \mu_i$, where $\sigma_i$ and $\mu_i$ are the standard deviation and mean of the time series, respectively. The bottom-left panel of Figure~\ref{fig:variability_beta2} shows the variability of $\angle\beta_2$, $M_{3,\,\angle\beta_2}$, plotted against the variability of the total intensity, $M_{3,\,I}$. 
At low mass accretion rates, cooling models exhibit substantially higher $\beta_2$ angle variability than non-cooling models. This is because the mean $\angle\beta_2$ in cooling cases is smaller while the standard deviation remains comparable, producing a larger $M_{3,\,\angle\beta_2}$ despite a similar absolute scatter. Interestingly, the intensity variability exhibits a bifurcation with observing frequency: cooling models show larger variability at 230~GHz but reduced variability at 86~GHz. The variability of $\angle\beta_2$  reflects intrinsic changes in the polarization structure. In cooling models, radiative feedback modifies the spatial and temporal distribution of polarised emission, leading to intermittent dominance of different emission regions and, consequently, rapid reorientation of the global EVPA pattern.

\subsection{Radius-dependent polarization structure}
\label{sec:radius-beta2}

\begin{figure*}
\centering
\includegraphics[width=2.0\columnwidth]{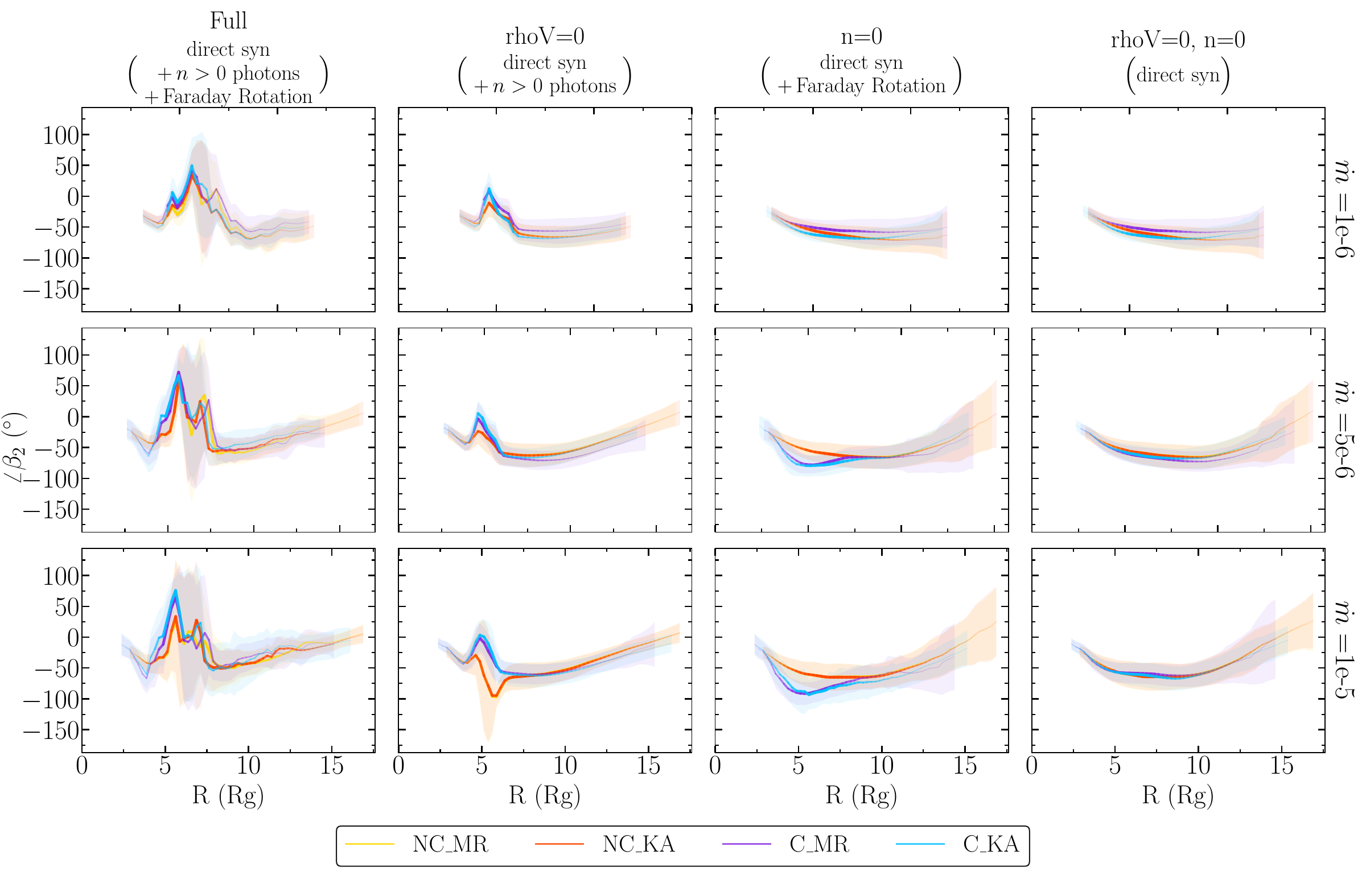}
\caption{Radial profile of polarization phase $\angle \beta_2$ for various GRMHD simulations at different mass accretion rates $\dot{m}$. In the second column, the Faraday effect is excluded for polarised light. In the third column, higher-order photons $n>0$ are excluded when computing the $\beta_2$ phase. In the fourth and last column, the polarization arises only from direct synchrotron emissions. From top to bottom, we show the cases at different mass accretion rates, $\dot{m}$ of $1 \times 10^{-6}$, $5 \times 10^{-6}$, $1 \times 10^{-5}$.  Only regions with $I > 0.1 \, I_{\rm max}$ are included. Line widths are weighted by intensity, and the shaded bands represent the variance of the $\angle \beta_2$ distribution at each radius.}
\label{fig:beta_R}
\end{figure*}

We further investigate the radial dependence of the quadrupolar polarization phase, $\angle\beta_2$, to disentangle the relative roles of spacetime effects and Faraday rotation in shaping the linear polarization structure. Figure~\ref{fig:beta_R} shows the radial profiles of $\angle\beta_2$ for different mass accretion rates and model variants, computed under controlled combinations of physical effects. 
To ensure robust polarization measurements, we exclude regions with intensities below $0.1\,I_{\rm max}$, where the polarization signatures become unreliable. The shaded bands indicate the variance of the $\angle \beta_2$ distribution in each radius, while the line widths are weighted by the local intensity contribution.

 Across all models and mass accretion rates, the polarization structure within $\sim5\,R_{\rm g}$, corresponding approximately to the region enclosed by the photon ring, remains remarkably similar regardless of whether Faraday rotation or higher-order photon trajectories are included. Moreover, the variance of $\angle \beta_2$ is generally small in this region, particularly at lower mass accretion rates where Faraday effects are weak. These results indicate that the polarization pattern near the black hole is primarily determined by strong-field gravitational effects and is largely insensitive to plasma thermodynamics or radiative-transfer propagation effects.

At larger radii ($R > 10\,R_{\rm g}$), the radial $\angle\beta_2$ profiles exhibit a systematic trend with accretion rate: $\angle\beta_2$ shifts from larger absolute values toward zero as $\dot{m}$ increases. This behaviour reflects intrinsic changes in the polarization pattern of the direct synchrotron emission, tracing the underlying plasma and magnetic-field structure in the outer disk, and persists even when additional geometric or propagation effects are included. Although $\angle\beta_2$ tends to become positive at very large radii ($R \gtrsim 15\,R_{\rm g}$) in models with higher accretion rate, this trend is not statistically significant given the substantially increased variance in the $\beta_2$ phase distribution at those radii.

A clear signature of higher-order photon contributions is revealed by comparing columns that include and exclude $n>0$ photons (first versus third columns, and second versus fourth columns). In these comparisons, we observe a sign reversal of $\angle\beta_2$ around $R\sim5\,R_{\rm g}$ when higher-order photons are included \citep{Chael2023ApJpolarization, Hou2025ApJ, Wong2026ApJ}. This flip arises from the photon-ring contribution, whose EVPA structure differs systematically from that of the direct emission. This behaviour is consistent with the sign change of $\angle\beta_2$ seen in the isolated $n=1$ images (see the Appendix, Figure~\ref{fig:Peak_pol}) and with previous analyses of photon-ring polarization \citep{Wong2026ApJ}. Importantly, when Faraday rotation is excluded (second column), the radial profiles at a given mass accretion rate are nearly identical across cooling models. However, we observe an opposite sign change towards negative values when the non-cooling models have larger mass accretion rates. This is puzzling, as the radius-integrated $n=1$ contributions show positive $\angle\beta_2$ for all the non-cooling models.

The role of Faraday rotation becomes apparent when comparing profiles with and without Faraday effects. At the lowest accretion rate, $\dot{m}=10^{-6}$, the radial profiles are nearly unchanged by the inclusion of Faraday rotation, especially when higher-order photons are not present. This confirms that Faraday rotation is negligible in the Faraday-thin regime. 
At higher accretion rates, $\dot{m}=5\times10^{-6}$ and $10^{-5}$, Faraday rotation modifies the radial structure of $\angle\beta_2$. 
Notably, when both Faraday rotation and higher-order photons are included, an additional flip in $\angle\beta_2$ appears just outside the photon ring region. This second flip does not occur when either effect is considered in isolation, demonstrating a non-trivial interplay between plasma propagation and spacetime-induced polarization transport. Interestingly, the location of the peak of the second flip differs between models with turbulence heating and reconnection heating, while it remains similar regardless of whether the cooling effect is included or not. This is consistent with the Faraday depth distributions shown in Fig.~\ref{fig:Hist_pol}, where $\tau_{\rho_V}$ is similar between non-cooling and cooling cases, but differs between the two heating prescriptions at the same mass accretion rate. For direct synchrotron emission, cooling models are especially sensitive to Faraday rotation, exhibiting a pronounced shift of $\angle\beta_2$ toward more negative values in the inner disk region ($5\,R_{\rm g} \lesssim R \lesssim 10\,R_{\rm g}$). This corresponds to a transition from a predominantly radial EVPA pattern toward a more circular configuration, driven by enhanced Faraday depth in simulations with radiative cooling.

\section{Summary and Discussions}
\label{sec:discussion}

In this section, we summarise and interpret the results presented in this paper, and discuss their observational relevance and the limitations of the numerical experiment. The simulations demonstrate that (i) radiative cooling and Coulomb coupling substantially modify the electron thermodynamic state and thereby change both intensity and polarization morphologies, (ii) Faraday propagation and strong-field (photon-ring) effects imprint distinct and, in some regimes, competing signatures on the linear polarization, and (iii) cooling tends to increase intrinsic asymmetry and temporal variability of the polarization even while it often reduces the effective Faraday scrambling at fixed accretion rate.

\subsection{Physical interpretation}

Radiative cooling and Coulomb coupling change the electron internal energy and the local emissivity distribution. In our models, this produces two principal, observationally relevant consequences. First, cooling reduces the electron temperature in regions where synchrotron and inverse-Compton losses are important, dimming and concentrating the disk emission relative to the jet/jet-sheath. Second, because Faraday rotation and Faraday depolarization depend sensitively on the electron density, temperature, and path length, cooling alters the local Faraday coefficients and hence the integrated Faraday depth along geodesics.

These changes yield a characteristic balance of effects. Cooling generally reduces large-scale Faraday scrambling in many regions (so that, at fixed $\dot{m}$, Faraday-induced changes in $|\beta_2|$ and $\angle\beta_2$ are smaller in cooling cases than in non-cooling cases), while simultaneously increasing intrinsic image asymmetry by concentrating emissivity into fewer, often transient, regions. The latter effect increases fractional power in asymmetric azimuthal modes ($m\neq2$), enhances ratios such as $|\beta_1|/|\beta_2|$, and produces larger temporal excursions and frequent phase reversals in $\angle\beta_2$.

The decomposition into emission order exposes the complementary role of strong-field gravitational transport. Photon-ring (higher-order) photons carry a polarization structure that can differ systematically from the direct (0th-order) emission; when the photon-ring fractional contribution to polarised flux is appreciable (enhanced in cooling cases where the direct disk emission is suppressed), it can produce sign flips in $\angle\beta_2$ at small radii ($\sim5\,R_{\rm g}$). Unlike Faraday effects, these strong-field imprints are comparatively insensitive to plasma thermodynamics and therefore represent a robust probe of relativistic polarization transport.

Finally, the radial dependence of $\angle\beta_2$ cleanly separates regimes: within $\sim5\,R_{\rm g}$ the polarization is dominated by spacetime/gravitational effects and is relatively robust to changes in electron physics; at larger radii plasma propagation (Faraday rotation/depolarization) and the spatial distribution of emissivity control the polarization morphology. When both Faraday rotation and higher-order photons act together, they can produce non-trivial additional flips in $\angle\beta_2$ whose radial location depends on the local Faraday depth.

\subsection{Observational implications and limitations}

Our results have direct implications for the interpretation of horizon-scale polarimetric measurements of systems such as M87* and Sgr~A* \citep{2019ApJ...875L...1E,2022ApJ...930L..12E}. In particular, the polarization morphology in the immediate vicinity of the black hole---including spiral or azimuthal EVPA patterns and photon-ring–induced flips in $\angle\beta_2$---is primarily determined by strong-field polarization transport and the large-scale magnetic geometry. These near-horizon signatures should therefore be relatively robust against uncertainties in electron microphysics.

 Photon-ring polarimetry may also encode additional information about black hole spin \citep{Himwich2020PhRvD} and turbulent variability \citep{Jimenez-Rosales2021MNRAS}. Moreover, \citet{Palumbo2022ApJ,Palumbo2023ApJ} showed that Faraday rotation leaves a characteristic imprint on the relation between direct and first-order polarization. Our results extend this picture by quantifying how Faraday depolarization affects higher-order polarization modes and by highlighting its potential to act as a confounding factor in spin inference, especially for future ngEHT measurements of the photon ring. 

At larger radii, however, the polarization structure becomes increasingly sensitive to plasma conditions. In this regime, the coherence of the quadrupolar mode $|\beta_2|$, the relative power in asymmetric modes, and the temporal variability of $\angle\beta_2$ all depend on the effective Faraday depth and on the extent to which radiative cooling redistributes the emissivity. Variations in these observables across epochs or frequencies may therefore reflect changes in $n_\mathrm{e}$, $B_\parallel$, or $T_\mathrm{e}$, rather than changes in the underlying magnetic topology alone. In particular, frequent EVPA reorientations and enhanced asymmetric-mode power favour regimes in which radiative feedback plays a significant role, making time-resolved and multi-epoch polarimetry especially valuable for constraining electron thermodynamics.

Quantitative comparison with EHT data will require realistic observational forward modelling, including finite angular resolution, sparse $uv$ coverage, thermal noise, calibration systematics, and, for Sgr~A*, interstellar scattering. Such modelling will be necessary before statistical measures such as $|\beta_2|$, $|\beta_1|/|\beta_2|$, and the radial dependence of $\angle\beta_2(R)$ can be robustly mapped to the underlying microphysical parameters.

\subsection{Limitations and outlook}

Several limitations constrain the generality of the present study. All simulations are performed in the magnetically arrested disc (MAD) state, characterised by strong, coherent magnetic flux; polarization properties in standard and normal evolution (SANE) discs may differ qualitatively. In addition, we consider a single black-hole spin, $a=0.9375$. Spin affects frame-dragging, photon-ring structure, and polarization transport, and the spacetime-induced polarization signatures highlighted here likely represent the stronger end of what would be expected across the full spin range.

The accretion rates explored, up to $\dot{m}=10^{-5}$ in Eddington units, emphasise radiative cooling and Faraday effects but may exceed those inferred for some low-luminosity sources. Consequently, the strength of Faraday-induced polarization signatures reported here should be interpreted with caution when compared to that of specific systems. Furthermore, we primarily adopt thermal electron distributions, with only a single exploratory non-thermal ($\kappa$) model. Real accretion flows likely host spatially and temporally varying non-thermal populations, and the electron heating prescriptions employed remain phenomenological approximations to collisionless kinetic physics. Finally, finite numerical resolution, limited temporal baselines, and the use of grid-scale dissipation models restrict our ability to capture all sources of variability and small-scale plasma effects.

Future work should extend this framework to include SANE accretion states, a broader range of black-hole spins, and more physically motivated, spatially varying electron distribution functions informed by kinetic calculations. Full forward-modelling of EHT observations, combined with multi-frequency and time-resolved polarimetry, will be essential for breaking degeneracies between spacetime effects and plasma microphysics. With these extensions, polarization diagnostics such as $\beta_2$, higher-order azimuthal modes, and their temporal and radial behaviour offer a promising pathway for constraining electron heating, cooling, and transport in accreting black holes.

\section{Conclusion}
\label{sec:conclusion}

We have investigated the polarization properties of near-horizon emission from radiatively inefficient accretion flows by combining two-temperature GRMHD simulations with polarised general relativistic radiative transfer. By systematically varying electron heating prescriptions, radiative cooling, and mass accretion rates in magnetically arrested disc (MAD) models, we have isolated how electron thermodynamics shapes horizon-scale polarization signatures.

Our results demonstrate that radiative cooling and Coulomb coupling play a central role in determining both the morphology and variability of polarised emission. Cooling modifies the spatial distribution of emissivity and the effective Faraday depth, leading to reduced large-scale Faraday scrambling at fixed accretion rate, while simultaneously enhancing intrinsic asymmetry and temporal variability in the polarization pattern. These effects are robust across different electron heating prescriptions and are most clearly captured by the behaviour of the quadrupolar polarization mode $\beta_2$ and the relative power in asymmetric azimuthal modes.

By decomposing the emission into photon-order and physical domain, we disentangle plasma propagation effects from strong-field gravitational polarization transport. We find that 
polarization within the photon-ring region ($\lesssim5\,R_{\rm g}$) is dominated by spacetime effects and is largely insensitive to electron thermodynamics, whereas
polarization at larger radii is controlled by Faraday rotation and the emissivity structure set by cooling. Photon-ring emission can induce characteristic sign reversals in $\angle\beta_2$, particularly when cooling suppresses the direct disk contribution, highlighting the importance of higher-order photons for interpreting horizon-scale polarimetry.

These results have direct implications for the EHT observations of sources such as M87* and Sgr~A*. While near-horizon polarization morphologies provide a robust probe of strong-field gravity and large-scale magnetic geometry, polarization amplitudes, asymmetric-mode power, and temporal EVPA variability offer sensitive diagnostics of electron thermodynamics and radiative feedback. Time-resolved and multi-frequency polarimetry, therefore, provides a promising pathway for constraining electron heating and cooling processes in accreting black holes.

\section*{Acknowledgements}
This research is supported by the National Key R\&D Program of China (grant no. 2023YFE0101200), the National Natural Science Foundation of China (grant No. 12273022, 12511540053), and the Shanghai municipality orientation program of basic research for international scientists (grant No. 22JC1410600).
JSL is supported by UCL through a UCL-RES scholarship and by the TDLI visiting student program. \\ 
MZ is supported by Doctoral Student Program of the Young S$\&$T Talents Cultivation Project, CAST and by T.D. Lee scholarship. \\
AU acknowledge the support from the Research Fund for Excellent International PhD Students (grant No. W2442004) \\  
YZ and KW are supported by a UKRI-STFC SA grant awarded to UCL-MSSL. 
JSL and KW acknowledge the support from the UCL Cosmoparticle Initiative. 
The simulations were performed on the TDLI-Astro cluster and the Siyuan Mark-I at Shanghai Jiao Tong University.

\section*{Data Availability}

The data underlying this article will be shared on reasonable request to the corresponding author.



\bibliographystyle{mnras}
\bibliography{example} 

\appendix

\section{Convolved image}
\label{app:blur}

\begin{figure*}
\centering
\includegraphics[width=2.0\columnwidth]{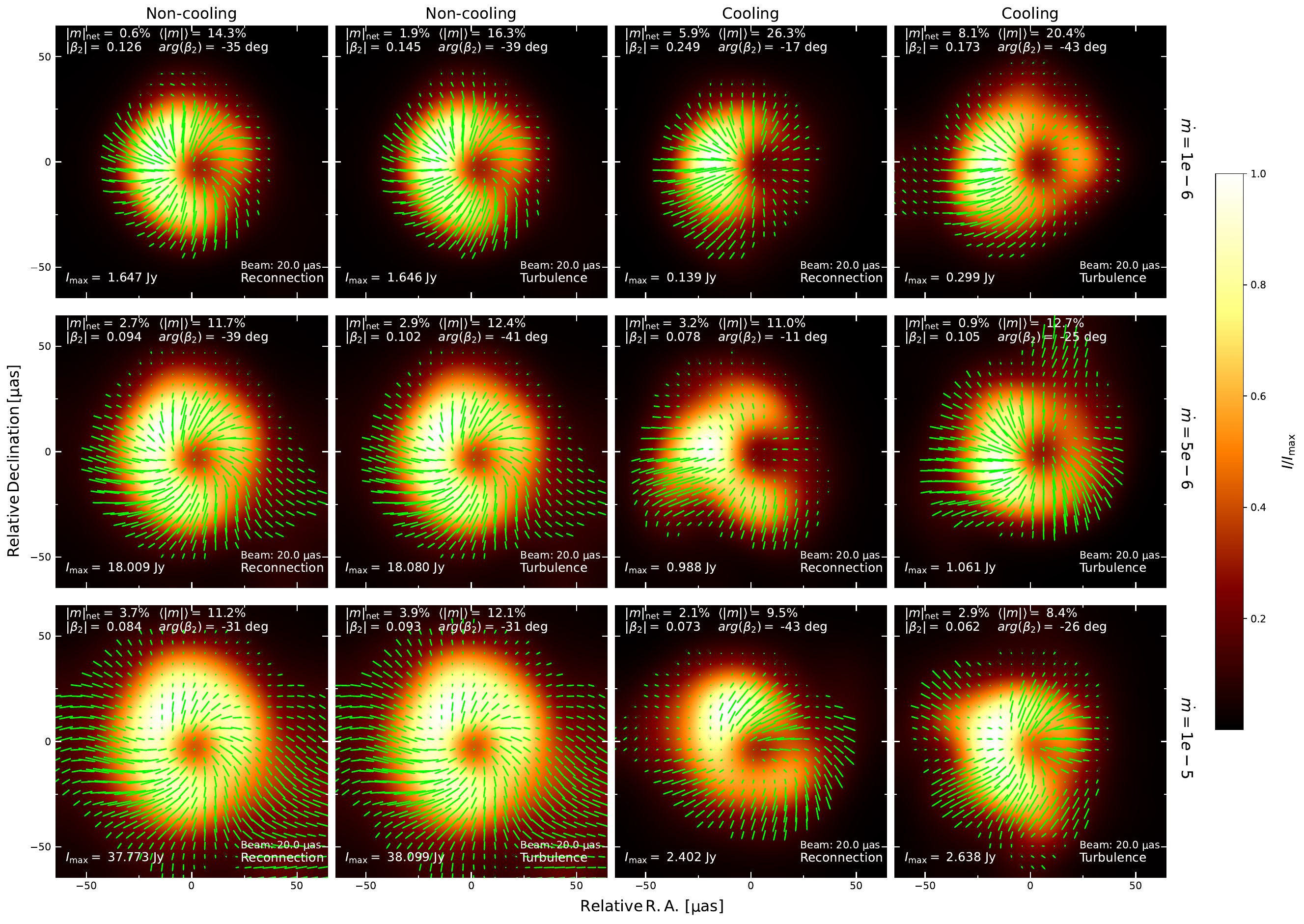}
\caption{Same as Fig.~\ref{fig:I}, except the panels show images with a $20 \mu as$ circular Gaussian blurring.}
\label{fig:blur}
\end{figure*}

To facilitate comparison with EHT observations, we apply a circular Gaussian convolution with a full width at half maximum of $20\,\mu\mathrm{as}$ to the images shown in Fig.~\ref{fig:I}. The resulting beam-convolved images are presented in Fig.~\ref{fig:blur}. While the total intensity morphology is smoothed on angular scales comparable to the EHT resolution, the large-scale EVPA structures show more coherent patterns. Notably, cooling models exhibit clear deviations from the nearly rotationally symmetric EVPA patterns seen in the non-cooling cases, indicating that the polarization morphology is sensitive to electron thermodynamics even after beam convolution, consistent with the stronger contributions of asymmetric modes shown in Fig~\ref{fig:beta_m_total}.

\section{Distributions of polarization properties}
\label{app:distribution}

\begin{figure*}
\centering
\includegraphics[width=1.0\columnwidth]{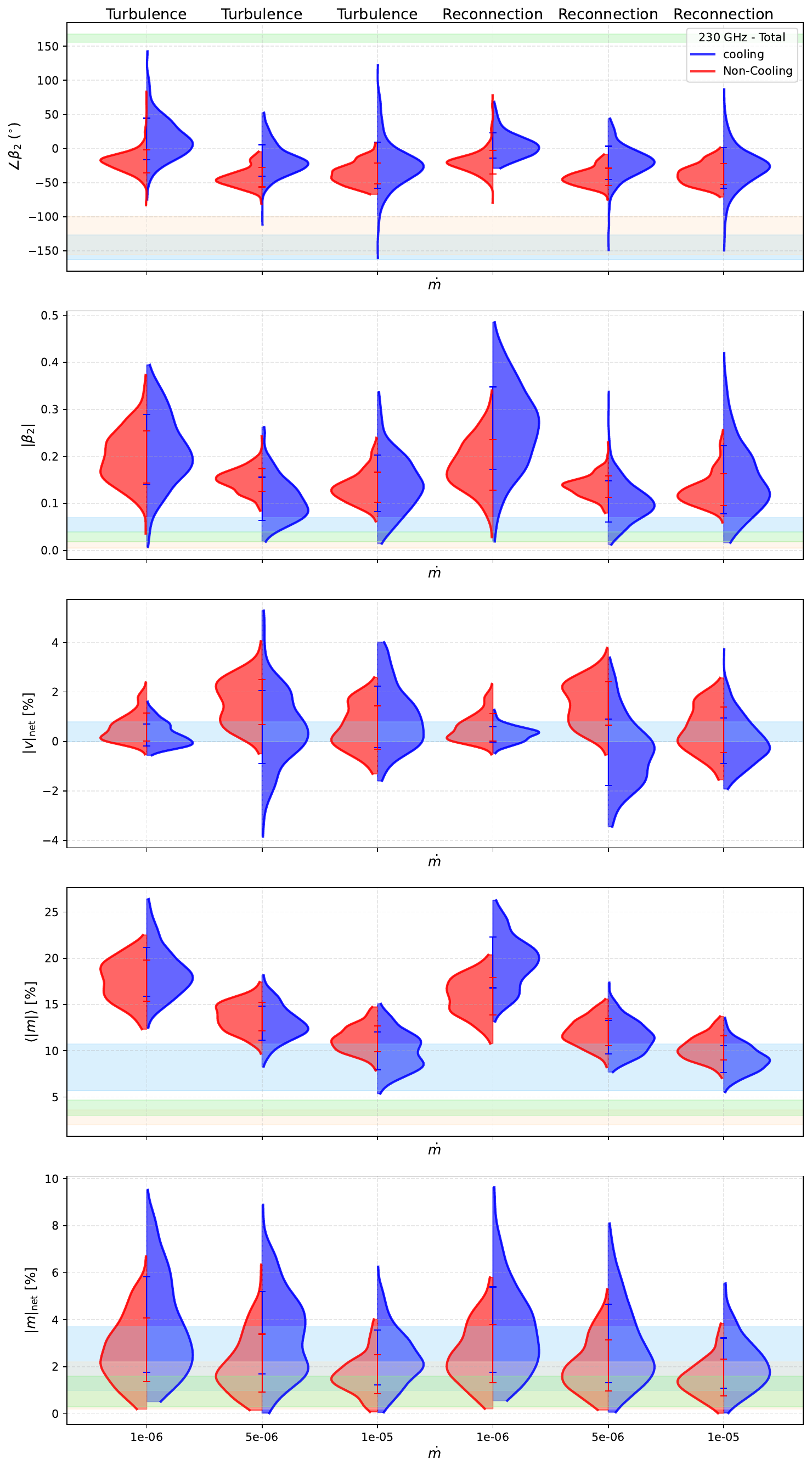}
\includegraphics[width=1.0\columnwidth]{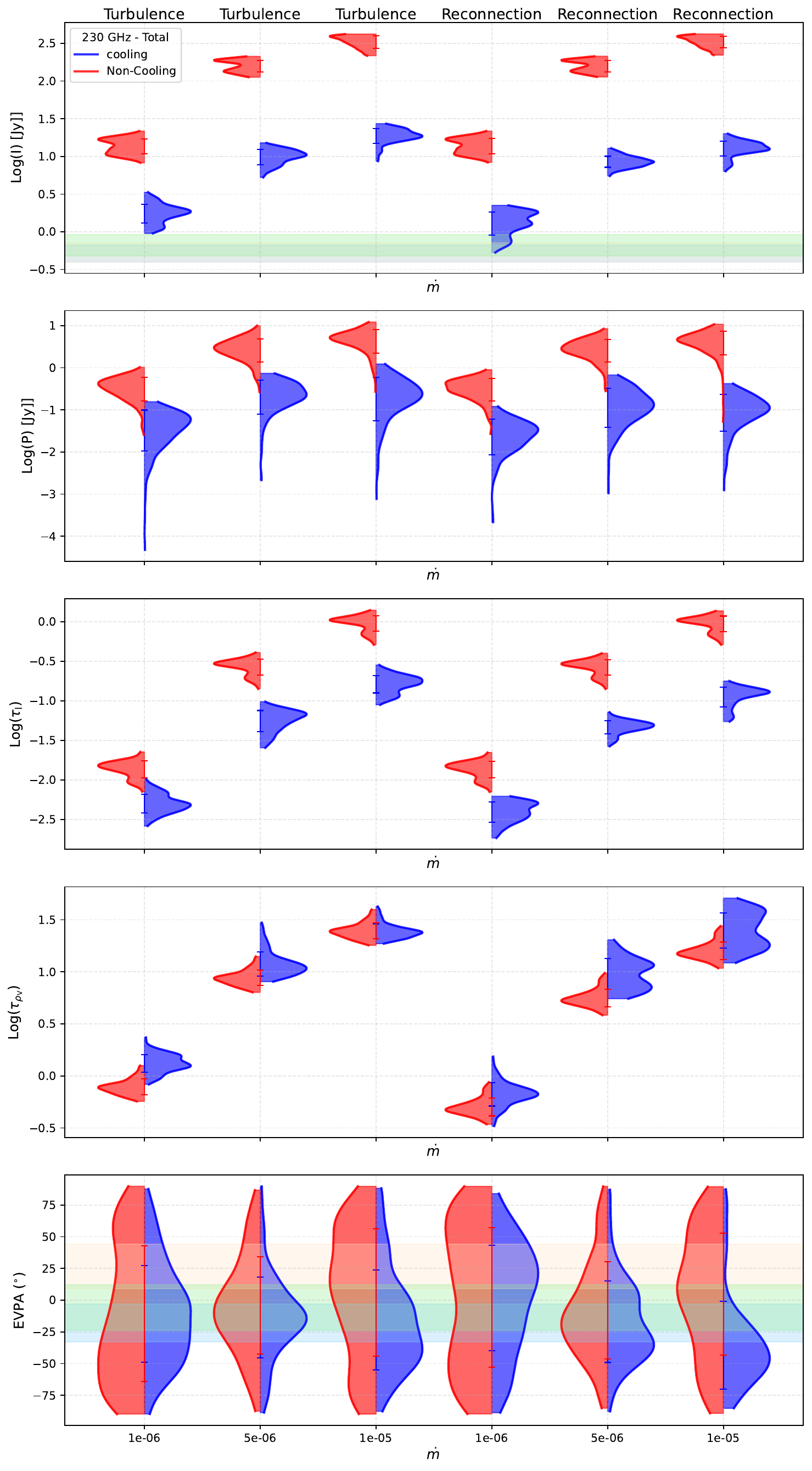}
\caption{Histogram of polarization properties at 230~GHz. The non-cooling and cooling cases are shown in red and blue, respectively. The left three columns are using turbulence with increasing mass accretion rates, while the right columns use reconnection heating.  The color of the bands (light-blue, light-orange, light-green) corresponds to different years of EHT observations of M87 in 2017, 2018, and 2021, respectively. }
\label{fig:Hist_pol}
\end{figure*}

We systematically analyse polarization properties across the full image ensemble and present their distributions as histograms in Fig.~\ref{fig:Hist_pol}. The quantities shown include the magnitude and phase of the second polarization mode, $|\beta_2|$ and $\angle\beta_2$; circular and linear polarization measures, $|v_{\rm net}|$, $\langle |m| \rangle$, and $|m_{\rm net}|$; and the image-integrated EVPA. In addition, we show the distributions of the total and polarised intensities ($I$ and $P$), as well as the optical depth $\tau_I$ and the Faraday rotation depth $\tau_{\rho_V}$. By comparing models with different electron thermodynamics and increasing mass accretion rates, these histograms allow us to identify systematic trends in polarization behaviour.


All models exhibit increasing depolarization with increasing mass accretion rate, as indicated by decreasing $|\beta_2|$ and $\langle |m| \rangle$. This behaviour is consistent with the rise in Faraday depth. Systems at low $\dot{m}$ typically have $\tau_{\rho_V} < 1$, implying weak Faraday rotation, whereas higher-$\dot{m}$ models reach $\tau_{\rho_V} \gtrsim 1$, leading to strong Faraday depolarization. Image-averaged linear polarization measures, $|\beta_2|$ and $\langle |m| \rangle$, are systematically lower in cooling models than in non-cooling models, although the image-integrated linear polarization fraction $|m_{\rm net}|$ shows comparatively little sensitivity to the cooling prescription.
At a fixed mass accretion rate, reconnection heating generally produces lower linear polarization than turbulence heating. In contrast, circular polarization $|v_{\rm net}|$ and the EVPA distribution show no big systematic differences across the models. Finally, both the total and polarised intensities are significantly reduced in cooling models relative to non-cooling models, consistent with their lower optical depths.

\begin{figure*}
\centering
\includegraphics[width=2.2\columnwidth]{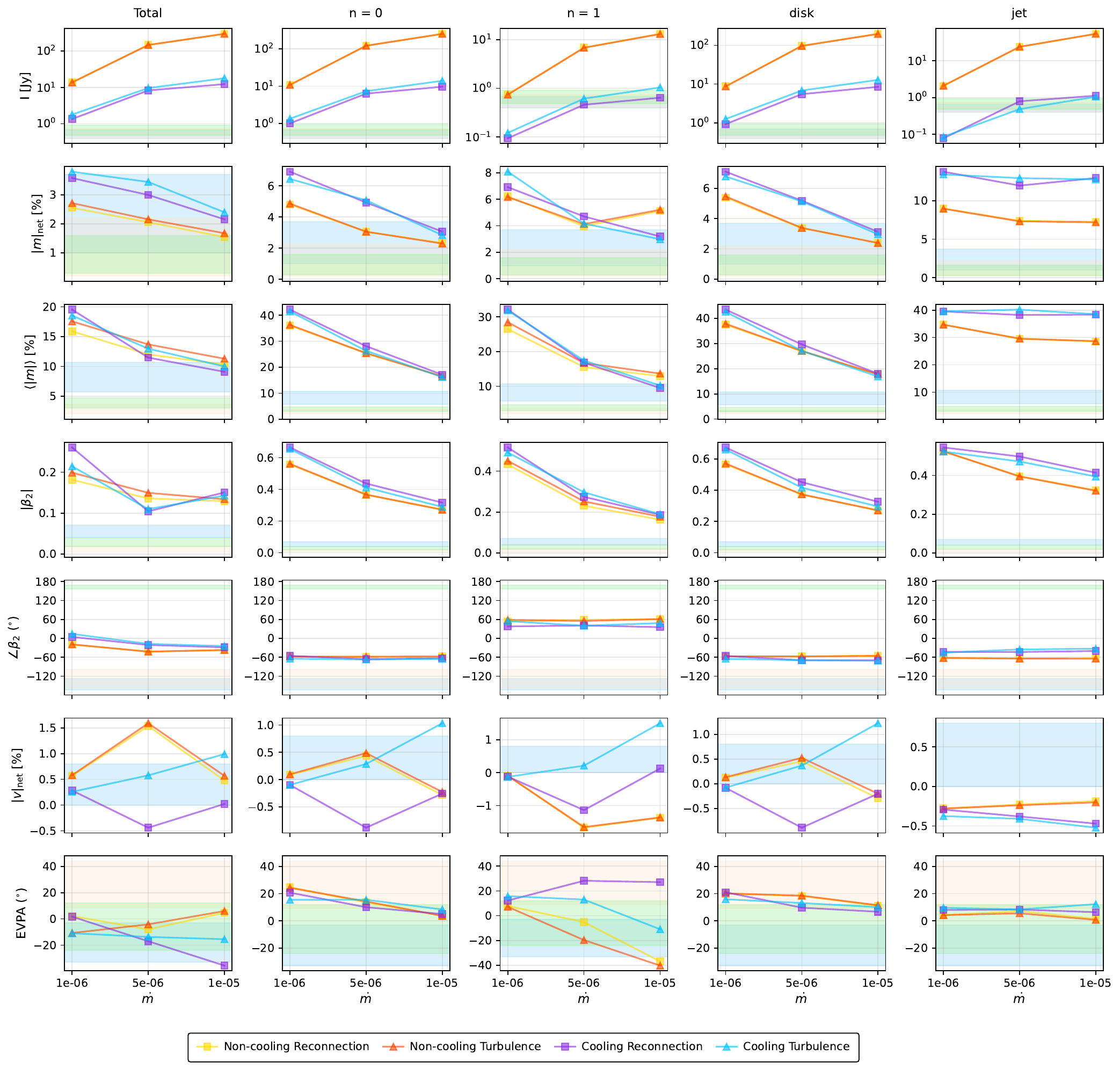}\caption{Variation of intensity $I$, net fractional polarization $|m|_{\rm net}$, resolved fractional polarization $\langle|m|\rangle$, amplitude and phase of the $\beta_2$ parameter, circular polarization $|v|_{\rm net}$, and the net EVPA for 230~GHz images with thermal synchrotron emissions. Various electron thermodynamics configurations are shown in legend, as a function of mass accretion rates $\dot{m}$. The values are computed as the mean of the histogram distribution of the GRMHD simulation spanning from $t=12000\,M$ to $15000\,M$.  Three color bands indicate constraints derived from the EHT observations of M~87 from 2017 (blue), 2018 (red), and 2021 (green) \citep{EHT2021ApJ1, EHT2021ApJ, EHT2025arXiv}. Each column represents a different emission region and photon orders. From left to right are the total region, $n=0$, $n=1$, and emissions from the disk and  jet of the $n=0$ images, respectively.
}
\label{fig:Peak_pol}
\end{figure*}

We summarise the ordering and domain-separated polarization properties for all GRMHD models in Fig.~\ref{fig:Peak_pol}. The quantities shown correspond to the mean values of the histogram distributions for each simulation. This approach provides a compact characterization of the dominant polarization properties in each model. Comparing the jet-only emission with that from the full GRMHD domain, we find that models employing turbulence heating systematically produce a larger fractional contribution from the jet than those using reconnection heating. For the three mass accretion rates considered without cooling, the jet emission fraction in turbulence-heated models is $5.0\%$, $5.5\%$, and $11.7\%$ at $\dot{m}=10^{-6}$, $5\times10^{-6}$, and $10^{-5}$, respectively. The corresponding fractions for reconnection-heated models are lower, at $3.9\%$, $4.7\%$, and $6.5\%$. This trend holds across frequencies and electron distribution prescriptions, indicating that turbulence heating enhances the relative importance of jet emission.

\begin{figure*}
\centering
\includegraphics[width=1.8\columnwidth]{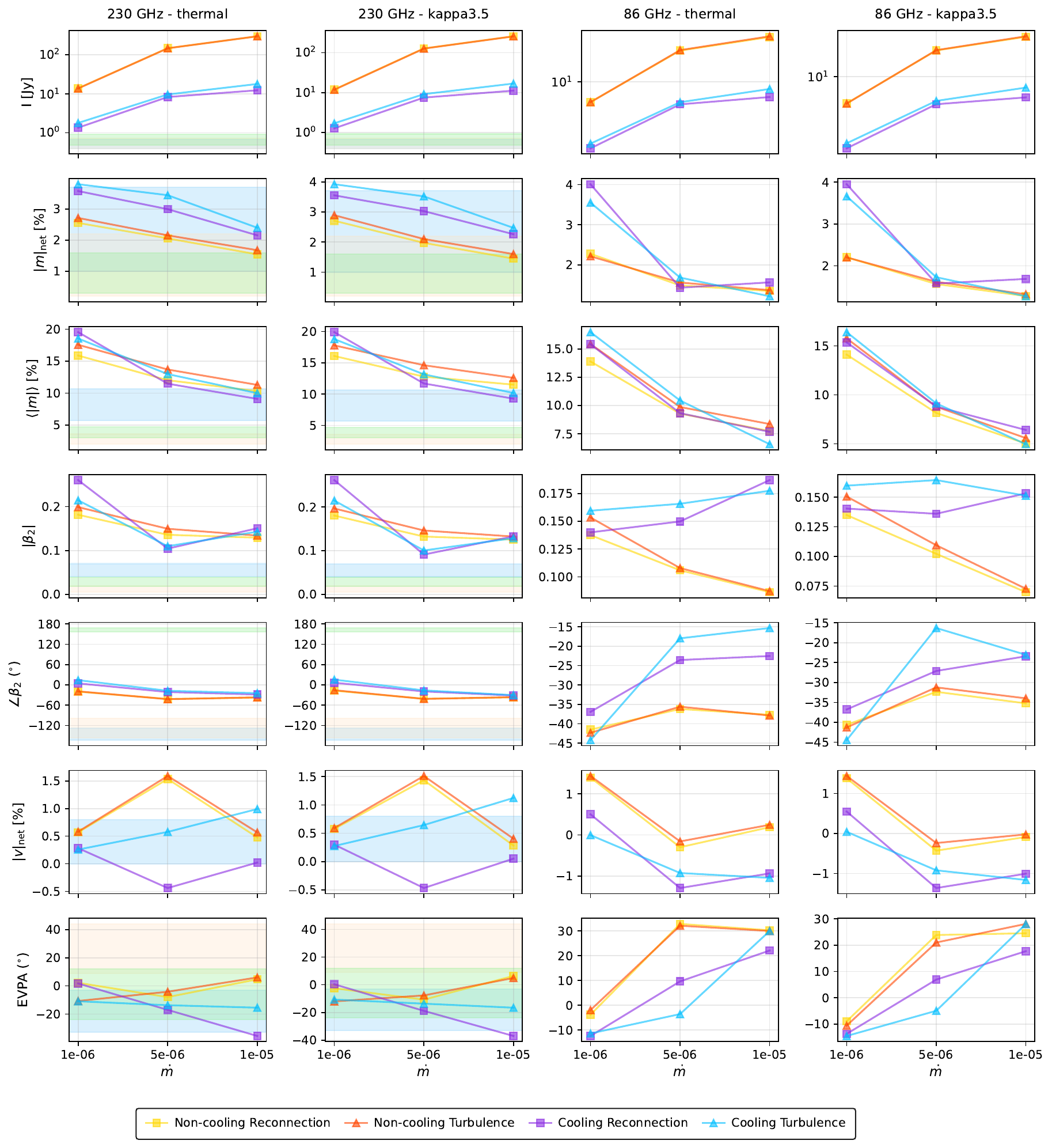}\caption{Same as Fig.~\ref{fig:Peak_pol}, except considering different electron density distributions and frequencies at 86 Hz. 
}
\label{fig:Peak_I}
\end{figure*}

Figure~\ref{fig:Peak_I} summarises the effect of non-thermal kappa electrons on  polarization quantities.   At 86 GHz, the amplitude and phase of $\beta_2$  change  slightly at higher $\dot{m}$ when kappa eDF is included. At 230 GHz, no visible differences are observed between thermal and non-thermal electron distributions.

\section{Light curve of the $\beta_2$ amplitude}
\label{app:beta_2}

\begin{figure*}
\vspace*{0.3cm}
\centering
\includegraphics[width=1.6\columnwidth]{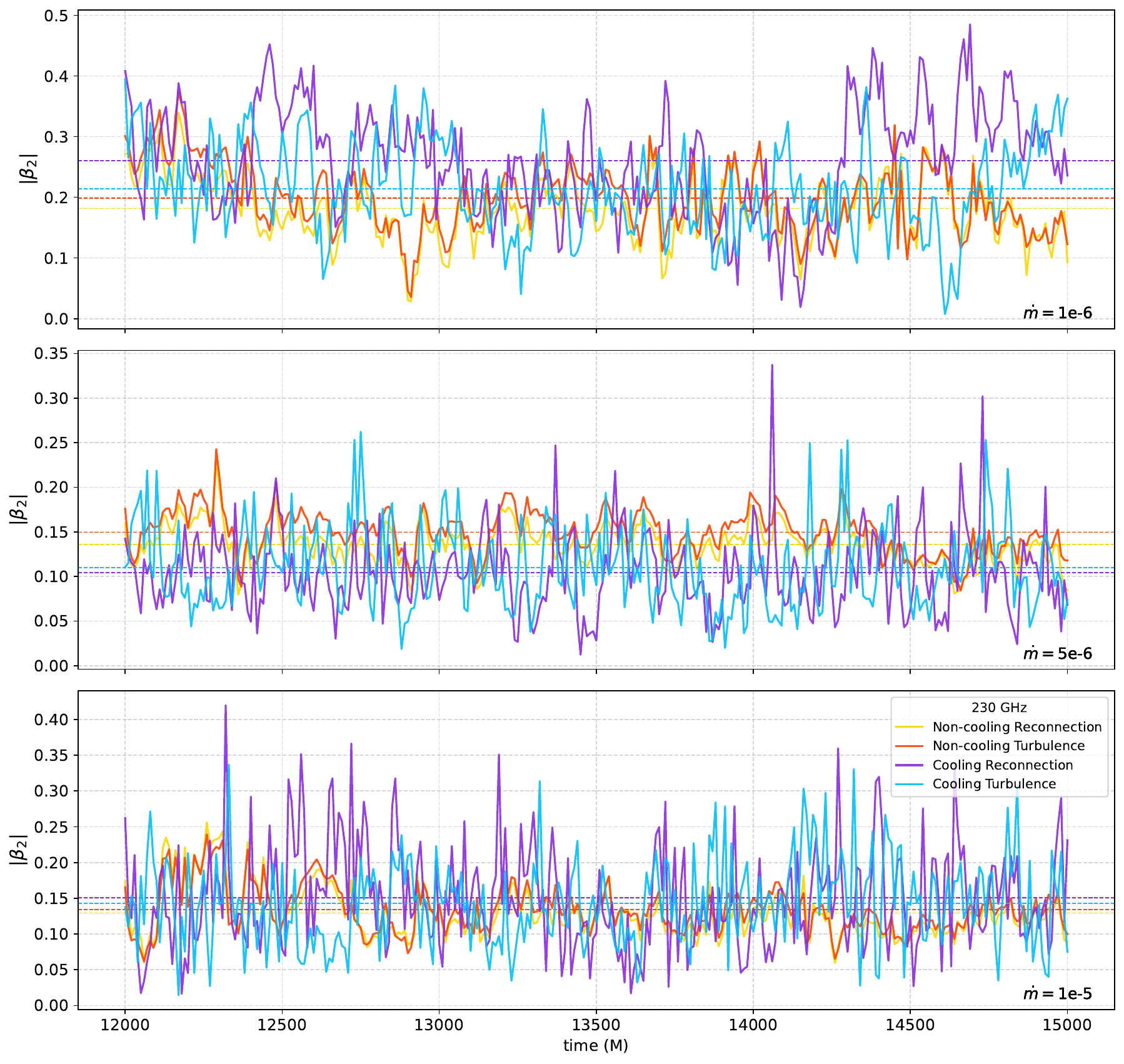}
\caption{Temporal evolutions of $|\beta_2|$.}
\label{fig:beta_2_lc}
\end{figure*}

 Figure~\ref{fig:beta_2_lc} shows the corresponding $|\beta_2|$ light curves discussed in Section~\ref{sec:beta2_var}. These light curves provide complementary information for interpreting the variability of $\angle\beta_2$. In particular, as mentioned in the main text, the phase reversals observed in the $\dot{m}=10^{-5}$ models near $t=12170\,M$ and $14780\,M$ coincide with exceptionally low values of $|\beta_2|<0.05$, indicating that the quadrupolar polarization mode is weak during these intervals. Consequently, the associated changes in $\angle\beta_2$ should be interpreted with caution, as they may reflect reduced polarization coherence rather than a robust reorientation of the global polarization pattern.

\bsp	
\label{lastpage}
\end{document}